\documentclass[aps,pre]{revtex4}
\usepackage{graphics}
\usepackage{graphicx}
\usepackage{amsfonts}
\usepackage{amssymb}
\usepackage{amsmath}
\usepackage{subfigure}
\usepackage{color}
\usepackage{natbib}

\begin{document}

\title{Cubic-Quintic Long-Range Interactions With Double Well Potentials}

\author{P.A.~Tsilifis}
\affiliation{Department of Mathematics, University of Southern
California, Los Angeles, 90089-2532 CA, USA}

\author{P.G. Kevrekidis}
\affiliation{Department of of Mathematics and Statistics, University of
Massachusetts,
Amherst, MA 01003-9305, USA}

\author{V.M.~Rothos}
\affiliation{Department of Mathematics, Physics Computational
Sciences, Faculty of Engineering, Aristotle University of Thessaloniki, Thessaloniki 54124, Greece}

\begin{abstract}
In the present work, we examine the combined effects of cubic
and quintic terms of the long range type in the dynamics of a double
well potential. Employing a two-mode approximation, we systematically
develop two cubic-quintic ordinary differential equations and assess
the contributions of the long-range interactions in each of the relevant
prefactors, gauging how to simplify the ensuing dynamical system.
Finally, we obtain a reduced canonical description for the conjugate
variables of relative population imbalance and relative phase between
the two wells and proceed to a dynamical systems analysis of the
resulting pair of ordinary differential equations. While in the
case of cubic and quintic interactions of the same kind (e.g. both
attractive or both repulsive), only a symmetry breaking bifurcation
can be identified, a remarkable effect that emerges e.g. in the setting
of repulsive cubic but attractive quintic interactions is a
``symmetry restoring'' bifurcation. Namely, in addition to the supercritical
pitchfork that leads to a spontaneous symmetry breaking of the
anti-symmetric state, there is a subcritical pitchfork that eventually
reunites the asymmetric daughter branch
with the anti-symmetric parent one.
The relevant bifurcations, the stability of the branches and their
dynamical implications are examined both in the reduced (ODE)
and in the  full (PDE) setting. The model is argued to be of
physical relevance, especially so in the context of optical
thermal media.
\end{abstract}

\maketitle

\section{Introduction}

In the study of both atomic and optical physics problems, often
analyzed in the realm of nonlinear Schr{\"o}dinger (NLS) type
equations~\cite{sulem,ablowitz}, the study of double well
potentials has a prominent position. Such potentials can be
straightforwardly realized in atomic Bose-Einstein condensates
(BECs) through the combination of a parabolic (harmonic) trap with
a periodic potential. Their experimental realization and
subsequent study in BECs with self-repulsive nonlinearity has led
to numerous interesting observations including tunneling and
Josephson oscillations for small numbers of atoms in the
condensate, and macroscopic quantum self-trapped states for large
atom number~\cite{markus1} and symmetry-breaking dynamical
instabilities~\cite{markus2}. These experimental developments have
been accompanied by a larger array of theoretical studies on
issues such as finite-mode reductions and symmetry-breaking
bifurcations
\cite{smerzi,kiv2,mahmud,bam,Bergeman_2mode,infeld,todd,theo},
quantum effects \cite{carr}, and nonlinear variants of the
potentials \cite{pseudo}. Similar features have also emerged in
nonlinear optical settings including the formation of asymmetric
states in dual-core fibers \cite{fibers}, self-guided laser beams
in Kerr media \cite{HaeltermannPRL02}, and optically-induced
dual-core waveguiding structures in photorefractive crystals
\cite{zhigang}.

On the other hand, a theme that has also been progressively
becoming of increasing importance within both of these areas
of physics is that of long range interactions. In the atomic
context, the experimental realization of BECs of magnetically polarized
$^{52}$%
Cr atoms \cite{Cr} (see recent review \cite{review} and for
a study of double well effects~\cite{pra09}),
as well as the study of dipolar molecules \cite%
{hetmol}, and atoms in which electric moments are induced by a
strong external field \cite{dc} have been at the center of the
effort to appreciate the role of long range effects. On the other
hand, in nonlinear optics, where nonlocal effects have been argued
to be relevant for some time now~\cite{krol1}, numerous striking
predictions and observations have arisen in the setting of thermal
nonlocal media~\cite{A}. Among them, we single out the existence of stable
vortex rings~\cite{krolik} the experimental realization of
elliptically shaped spatial solitons~\cite{moti1} and the
observation of potentially pairwise attracting (instead of
repelling as in the standard local cubic media) dark
solitons~\cite{krolik2}. Another very important large class of
systems displaying a nonlocal nonlinearity consists of materials with a
quadratic nonlinearity. In ~\cite{B}, it has been
shown that, in fact, the quadratic
nonlinearity is inherently nonlocal.  This implies that nonlocality
can be used explain the beautiful X-wave~\cite{C} observations
and even the different
regimes of soliton pulse compression in quadratic
materials~\cite{D,E}. It is interesting to note that in these
quadratic media, not only does the prototypical
ingredient of (effective) nonlocality arise, but it is also possible
for a competition of this feature with the cubic nonlinearity to emerge 
as is discussed in ~\cite{F}.

Our aim in the present work is to expand on the framework of
studies of double well potentials in the presence of nonlocal
nonlinear interactions by considering cubic-quintic models. Part
of the motivation for doing so consists of the fundamental
relevance of the cubic-quintic NLS. The latter is a model that has
been used in a variety of physical settings. These include the
light propagation in optical media such as non-Kerr crystals
\cite{PTS}, chalcogenide glasses \cite{CQglass}, organic materials
\cite{CQorganic}, colloids \cite{colloid}, dye solutions
\cite{dye}, and ferroelectrics \cite{ferroelectric}. It has also
been predicted that this type of nonlinearity may be synthesized
by means of a cascading mechanism \cite{cascading}. An additional
part of the motivation stems from an interesting set of
observations that were made in an earlier work featuring competing
{\it cubic} nonlinearities, one of which was a cubic local and
another was a cubic nonlocal one; see~\cite{chenyu} and the
discussion therein. In that work, it was found that for repulsive
nonlocal cubic interactions and attractive local ones, it was
possible to tune the prefactors determining the competition so as
to produce not only a symmetry breaking, but also a
symmetry-restoring bifurcation. More recently, a similar
conclusion in a local cubic-quintic double well potential was
reached in~\cite{jy}.

Here, we present a framework where the competition of cubic and
quintic terms can be systematically quantified. In addition, to
address the problem from a broader perspective, we consider fully
nonlocal interactions both for the cubic and the quintic terms,
rendering the local case a straightforward special-case scenario
of our study. The specific setup we consider here is partially
of interest to the field of cold gases e.g. in the case
of repulsive quintic (but local) interactions and attractive
cubic nonlocal ones. This exactly corresponds to the
model of the dipolar Tonks-Girardeau gas with the dipole moments
polarized along the axis, considered earlier in~\cite{boris_add}.
The difference here is that in this setting the quintic interaction
cannot be made nonlocal (although the relevant mathematical norm
form description and physical phenomenology will be essentially the same as 
presented herein). A setup more precisely in tune with the considerations
given below arises in the field of nonlinear optics and, more
particularly, in the case of thermal optical nonlinearity~\cite{A,boris_add2} 
but when the heating is
provided by the resonant absorption by dopants,  in which case
the absorption may be saturable. In the appendix, we justify more
precisely this connection to the specific model 
analyzed in what follows.

We start our presentation of the theoretical
analysis of section II by developing a two-mode
reduction of the system with both the cubic and the quintic terms.
We systematically examine all the relevant terms and offer a
prescription for assessing the dominant contributions
to the resulting dynamics of the left and the right well.
Following an amplitude-phase decomposition and examining the
variables associated with the population imbalance of the
two wells, and their relative phase, we construct the
Hamiltonian normal form of the two-mode reduction of the cubic-quintic
double well system. We then {\it explicitly illustrate} how the
bifurcation analysis of this normal form encapsulates not only
the symmetry breaking but {\it also} the symmetry restoring.
We argue that this cubic-quintic realization is the prototypical
one where both of these effects can be observed and analytically
quantified. Subsequently, in section III, we proceed to test
the relevant predictions by means of a computational bifurcation
analysis, as well as through direct numerical simulations (in order
to monitor the predicted dynamical instabilities).
We find very good agreement with the symmetry breaking predictions
of the model and even a quite fair agreement with the symmetry
restoring ones (which arise in a highly nonlinear regime and
are hence less amenable to a two-mode analysis). We also quantify
the disparity of the analytical predictions and numerical results
for large values of the nonlocality range parameter. Finally,
section IV contains our conclusions and some directions for
future study.

\section{Analytical approach for the NLS equation
  with two nonlocal terms}

\subsection{Two-mode approximation}

As indicated above, our fundamental model will be the 1d NLS equation
in the presence of two nonlocal terms, namely the cubic and quintic ones:
\begin{eqnarray}
i\partial_t\psi + \mu\psi = {\cal L}\psi +
s\left(\int_{-\infty}^{+\infty} R_1(x - x')|\psi(x')|^2dx'\right)\psi +
\delta\left(\int_{-\infty}^{+\infty}R_2(x - x')|\psi(x')|^4dx'\right)\psi
\label{eq1}
\end{eqnarray}
with $s,\delta = \pm1$ and the linear operator will be of the
standard Schr{\"o}dinger type
\begin{eqnarray*}
{\cal L}\  = -(1/2)\partial^2_x + V(x).
\end{eqnarray*}
This encompasses the
double-well potential of the form:
\begin{eqnarray*}
V(x) = (1/2)\hat{\Omega}^2 x^2 + V_0\rm sech^2(x/w)
\end{eqnarray*}
with $\hat{\Omega}$ being the normalized strength of the parabolic
trap and it is
$\hat{\Omega} \ll 1 $ in a quasi-1d situation in BECs (here
the effective trap frequency is the ratio of the longitudinal trap
strength along the condensate over the one of the
tightly confined transverse directions). In our study we consider a
typical experimentally relevant value of $\hat{\Omega} = 0.1$,
while the generally tunable (see e.g.~\cite{engels}) parameters
of the laser beam forming the light defect
are chosen to be $V_0 = 1$ and $w = 0.5$ (which we have found to be
fairly typical values representative of the phenomenology to be
analyzed below).

For the kernels $R_1$, $R_2$ we will focus our considerations on
either the Gaussian
\begin{eqnarray*}
R_i(x) = \frac{1}{\sigma\sqrt{\pi}}\exp(-\frac{x^2}{\sigma^2})
\end{eqnarray*}
or the exponential
\begin{eqnarray*}
R_i(x) = \frac{1}{2\sigma}\exp(-\frac{|x|}{\sigma}).
\end{eqnarray*}
While the latter is more specifically relevant to the 
thermal nonlocal (optical) media and to quadratic nonlinear
materials~\cite{krol1,B,D,E}, we also use the former due to the 
mathematical simplicity of its kernel. In any event, our results
will not be significantly different qualitatively between the two
cases, although obviously the quantitative details will not be the same.
The key parameter
here is the range of the nonlocal interaction parametrized by
$\sigma$. Notice that both kernels in the limit of
$\sigma \rightarrow 0$ tend to a genuinely local interaction
(i.e., $R_i(x) \rightarrow \delta(x)$).

\begin{figure}[th]
\begin{center}
\includegraphics[width = 9cm]{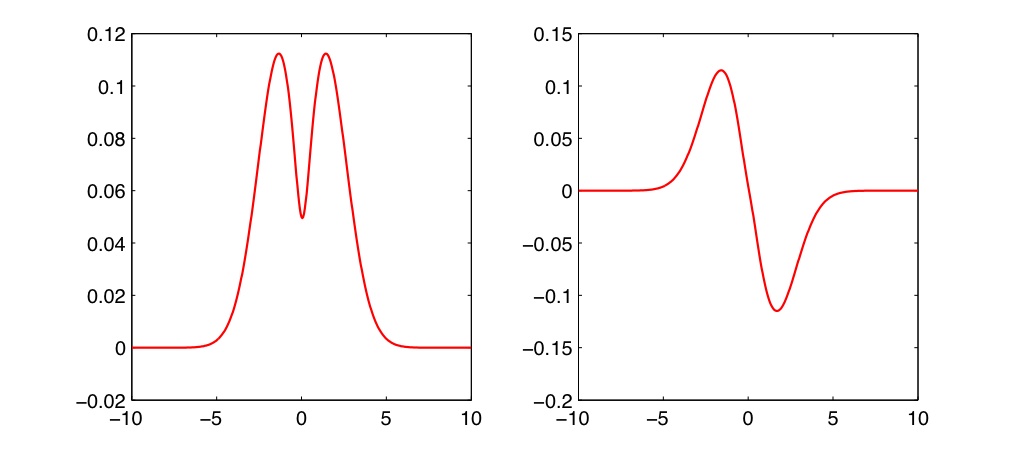}
\includegraphics[width = 9cm]{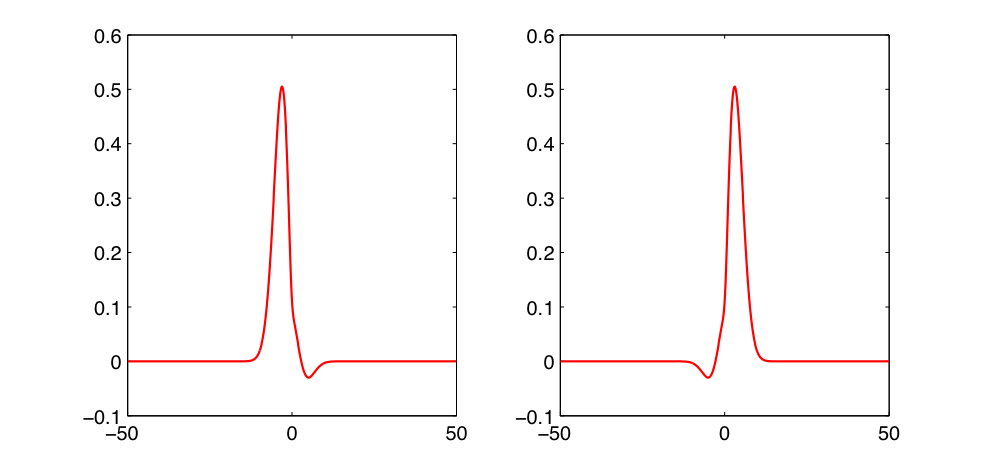}
\end{center}
\caption{ The ground state $u_0$ and first excited state $u_{1}$ of the
potential are shown in the top panel. The rotated
orthonormal basis of $\phi_L$ and $\phi_R$ (with support, respectively,
on the left and right well) is shown in the bottom panels.}
\label{basis}
\end{figure}

We now develop the two-mode approximation in order to
obtain a decomposition (or more accurately a Galerkin truncation)
of the solution $\psi$ over the minimal
basis of fundamental states. More specifically,
we use an orthonormal basis composed by the wave functions
$\{\phi_L,\phi_R\} \equiv \{(u_0-u_1)/\sqrt{2},(u_0+u_1)/\sqrt{2}\}$, where $u_0$ and $u_1$ (Fig. 1) are the ground state and the first excited state,
respectively,
corresponding to the first two eigenvalues of ${\cal L}$ that are
$\omega_0 = 0.13282$ and $\omega_1 = 0.15571$ for our choice
of potential parameters above.
Notice that these two eigenfunctions
represent modes with support predominantly on the left and right well,
respectively. The eigenfunctions $u_{0,1}$ and the rotated basis employed
herein of $\phi_{L,R}$ are both shown in Fig.~\ref{basis}.

The two-mode approximation is then defined as
\begin{eqnarray}
\psi(x,t) = c_L(t)\phi_L(x) + c_R(t)\phi_R(x)
\label{twomode}
\end{eqnarray}
where $c_L$ and $c_R$ are complex time-dependent amplitudes and
the approximation consists of the truncation of the higher modes
within the expansion. Before substituting into the initial
Gross-Pitaevskii (GP) equation, we notice that the action of the
linear operator ${\cal L}$ on our basis elements is as follows:
\begin{eqnarray*}
{\cal L} \psi = (\Omega c_L - \omega c_R)\phi_L + (\Omega c_R -
\omega c_L)\phi_R
\end{eqnarray*}
where $\Omega = (\omega_0 + \omega_1)/2$ and $\omega = (\omega_1 -
\omega_0)/2$ are linear combinations of the two eigevalues of
${\cal L}$ respectively to the solutions $u_0$, $u_1$. Subsequently,
substitution of our ansatz of Eq.~(\ref{twomode}) in the
full nonlinear problem of Eq.~(\ref{eq1}) yields:
\begin{eqnarray*}
i\dot{c}_L\phi_L + i\dot{c}_R\phi_R = (\Omega c_L - \mu c_L - \omega
c_R)\phi_L + (\Omega c_R - \mu c_R - \omega c_L)\phi_R +\\
+ s|c_L|^2(c_L\phi_L + c_R\phi_R)\int R_1(x - x')\phi_L^2(x')dx' +
s|c_R|^2(c_L\phi_L + c_R\phi_R)\int R_2(x - x')\phi_R^2(x')dx' \\
+ s[(c_L^2c_R^*+ |c_L|^2c_R)\phi_L + (c_L^*c_R^2 +
c_L|c_R|^2)\phi_R]\int R(x - x')\phi_L(x')\phi_R(x')dx'\\
+ \delta|c_L|^4(c_L\phi_L + c_R\phi_R)\int R_2(x - x')\phi_L^4(x')dx'
+ \delta|c_R|^4(c_L\phi_L + c_R\phi_R)\int R_2(x -
x')\phi_R^4(x')dx'\\
+ \delta\left[(4|c_L|^4|c_R|^4c_L + c_L^3{c_R^*}^2 + c_L^*|c_L|^2c_R^2)\phi_L
+ (4|c_L|^4|c_R|^4c_R + c_R^3{c_L^*}^2 + c_R^*|c_R|^2c_L^2)\phi_R
\right] \cdot\\ \cdot \int R_2(x - x')\phi_L^2(x')\phi_R^2(x')dx'\\
+ 2\delta\left[(|c_L|^2c_L^2c_R^* + |c_L|^4c_R)\phi_L +
  (|c_L|^2|c_R|^2c_L + |c_L|^2c_R^2c_L^*)\phi_R\right]\int R_2(x -
x')\phi_L^3(x')\phi_R(x')dx'\\
+ 2\delta\left[(c_L^2|c_R|^2c_R^* + |c_L|^2|c_R|^2c_R)\phi_L +
  (|c_R|^4c_L + c_R^2|c_R|^2c_L^*)\phi_R \right]\int R_2(x -
x')\phi_R^3(x')\phi_L(x')dx'.
\end{eqnarray*}

\begin{figure}[th]
\begin{center}
\includegraphics[width = 8.1 cm]{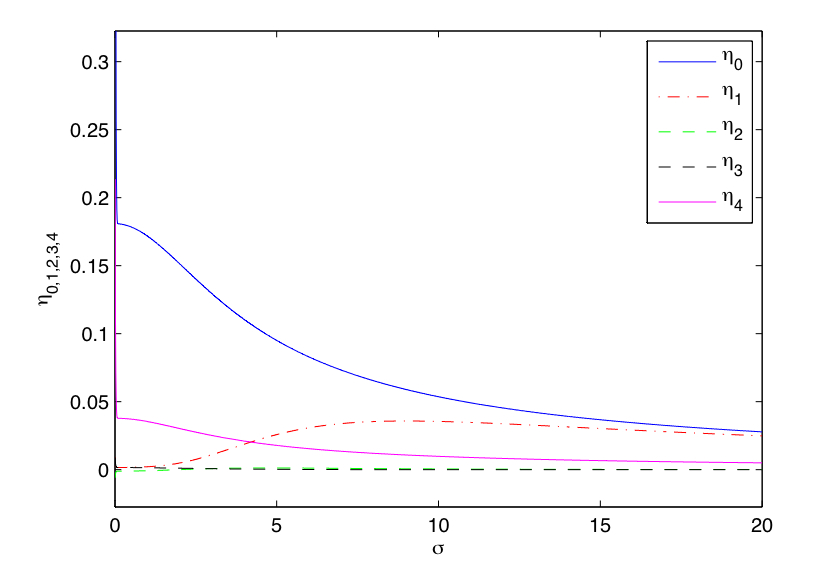}
\includegraphics[width = 8cm]{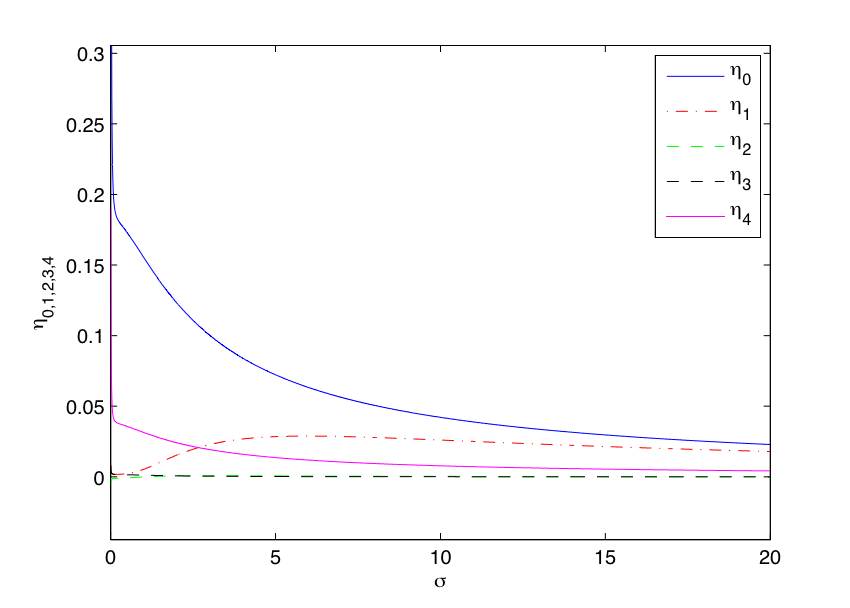}
\end{center}
\caption{The overlap integrals $\eta_0$, $\eta_1$, $\eta_2$, $\eta_3$ and $\eta_4$ are shown as a function of the interaction range
$\sigma$ for the Gaussian (left) and exponential (right) kernels.}
\label{overlap_int}
\end{figure}

In order to project the above equation onto the states $\phi_{L,R}$
we multiply with the respective function (notice that
the eigenfunctions are real due to the Hermitian nature
of the operator ${\cal L}$) and integrate. This involves
the following integrals which will play a fundamental role
in our considerations below:
\begin{eqnarray*}
\eta_0 = \int\int R_1(x - x')\phi_L^2(x')\phi_L^2(x)dx'dx,\\
\eta_1 = \int\int R_1(x - x')\phi_L^2(x')\phi_R^2(x)dx'dx,\\
\eta_2 = \int\int R_1(x - x')\phi_L^2(x')\phi_L(x)\phi_R(x)dx'dx,\\
\eta_3 = \int\int R_1(x - x')\phi_L(x')\phi_R(x')\phi_L(x)\phi_R(x)dx'dx,\\
\end{eqnarray*}
from the first nonlocal term, as well as
\begin{eqnarray*}
\begin{array}{cc} \eta_4 = \int\int R_2(x - x')\phi_L^4(x')\phi_L^2(x)dx'dx, & \eta_8 = \int\int R_2(x - x')\phi_L^2(x')\phi_R^2(x')\phi_L(x)\phi_R(x)dx'dx,\\
 & \\
\eta_5 = \int\int R_2(x - x')\phi_L^4(x')\phi_R^2(x)dx'dx, & \eta_{9} = \int\int R_2(x - x')\phi_L^3(x')\phi_R(x')\phi_L^2(x)dx'dx, \\
 & \\
\eta_6 = \int\int R_2(x - x')\phi_L^4(x')\phi_L(x)\phi_R(x)dx'dx, & \eta_{10} = \int\int R_2(x - x')\phi_L^3(x')\phi_R(x')\phi_R^2(x)dx'dx,\\
 & \\
\eta_7 = \int\int R_2(x - x')\phi_L^2(x')\phi_R^2(x')\phi_L^2(x)dx'dx, & \eta_{11} = \int\int R_2(x - x')\phi_L^3(x')\phi_R(x')\phi_L(x)\phi_R(x)dx'dx\\
 & \\
\end{array}
\end{eqnarray*}
from the second nonlocal term. Some alternatives that are derived if
we interchange the variables $x$ and $x'$ or swap $L$ and $R$ can also
be equivalently considered. A numerical study of the first four intergrals
was already conducted in~\cite{chenyu}, where it was found that typically
the integrals $\eta_{2,3}$ can be considered as negligible in
comparison to $\eta_0$ which is the  dominant term. On the
other hand, $\eta_1$ is close to $\eta_{2,3}$ for near-local
interactions (i.e., for small values of
$\sigma$), but becomes comparable to $\eta_0$ as the latter descreases
for wide nonlocal interaction ranges (i.e., for large $\sigma$). The criterion
that we use to determine whether $\eta_1$ is negligible or not was
$\eta_{rel} \geq 0.01$ where $\eta_{rel} = \eta_1 - \max (|\eta_2|,|\eta_3|) $.
This yields that $\eta_1$ remains significant until (i.e., down to)
a critical value $\sigma_b = 2.96$ and $1.56$ for the Gaussian and
exponential kernels, respectively. The dependence of the
relevant overlap integrals on the range of the interaction
$\sigma$ is shown in Fig.~\ref{overlap_int}.

\begin{figure}[th]
\begin{center}
\includegraphics[width = 8cm]{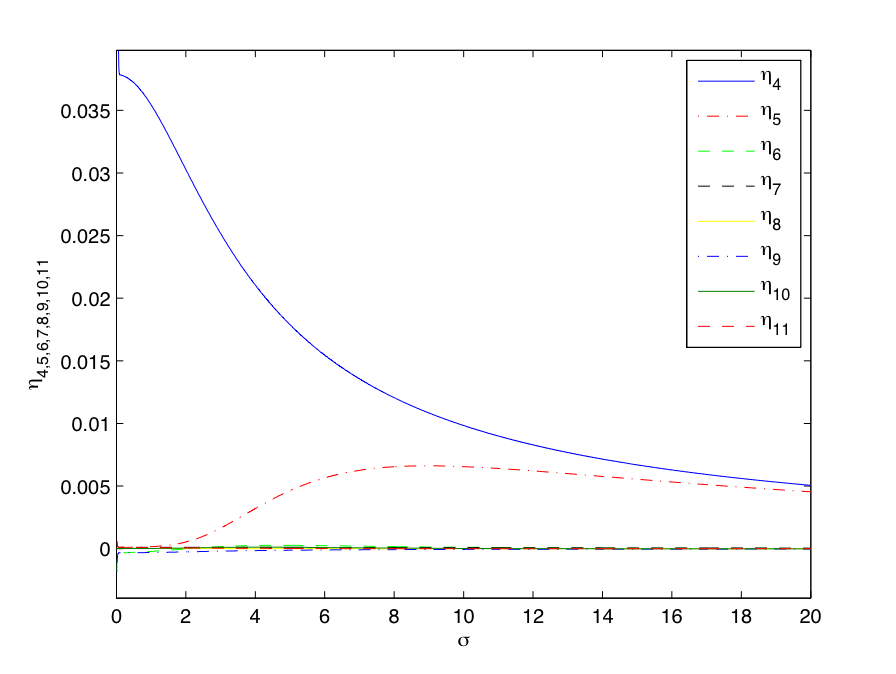}
\includegraphics[width = 8cm]{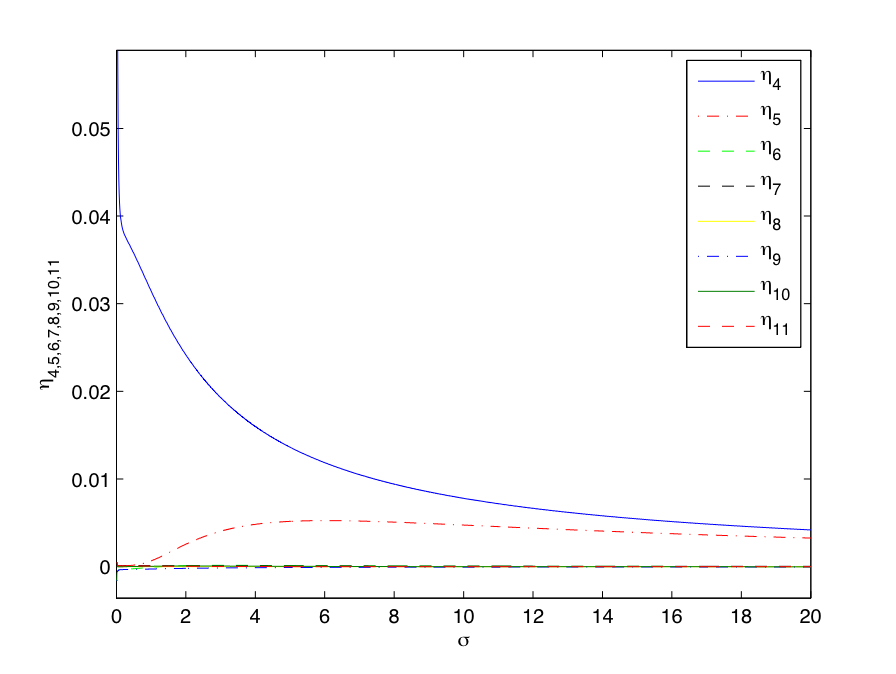}
\end{center}
\caption{The overlap integrals $\eta_{4,5,\dots,11}$ are given here
as a function of the interaction range $\sigma$, for the two kernels
in order to appreciate the dominance of $\eta_4$ with respect to the
remaining terms for the range $\sigma < \sigma_c$, where the term
with prefactor $\eta_4$ is not negligible with respect to the overall
dominant term $\eta_0$.}
\label{overlap_int2}
\end{figure}

Taking into regard the second nonlocal term (which for simplicity we have
assumed to share the same range parameter as the first),
we can see from Fig.~\ref{overlap_int2} that the integrals $\eta_{5,6,...}$ are
always negligible but $\eta_{4}$ appears to be a nontrivial
competing term. This is to a certain degree intuitively
anticipated, as this represents
the dominant term associated with the quintic interaction.
Adapting the same criterion as in~\cite{chenyu} (namely $\eta_{rel} = \eta_4 - \max (|\eta_2|,|\eta_3|)$), we incorporate the relevant $\eta_4$
for $\sigma < \sigma_{c} = 9.15$, $7.01$ for the Gaussian and exponential
kernel, respectively. According to this we may distinguish three cases:
\begin{itemize}
\item The terms $\eta_0$ and $\eta_4$ are considered for $\sigma < \sigma_b = 2.96$ (for the Gaussian kernel);
\item The term with prefactor
$\eta_1$ is added when $\sigma_b < \sigma < \sigma_c$.
\item For $\sigma > \sigma_c$, $\eta_4$ is omitted and only
$\eta_0$, $\eta_1$ are taken into account.
\end{itemize}

For the first case, the projection of the equation onto the states $\phi_{L,R}$ yields
\begin{eqnarray*}
i\dot{c}_L = (\Omega - \mu)c_L - \omega c_R + s\eta_0|c_L|^2c_L + \delta\eta_4|c_L|^4c_L\\
i\dot{c}_R = (\Omega - \mu)c_R - \omega c_L + s\eta_0|c_R|^2c_R + \delta\eta_4|c_R|^4c_R,
\end{eqnarray*}
and by introducing Madelung representation of
action-angle or amplitude-phase decomposition
($c_{L,R} = \rho_{L,R}e^{i\theta_{L,R}}$), we obtain
\begin{eqnarray}
\left\{ \begin{array}{cc} \dot{\rho}_L = \omega\rho_R\sin\theta\hspace{3.5 cm}\\ \\
\dot{\theta}_L = \mu - \Omega + \omega\frac{\rho_R}{\rho_L}\cos\theta - s\eta_0\rho_L^2 - \delta\eta_4\rho_L^4, \end{array} \right\}
\end{eqnarray}
where we have defined the relative phase
$\theta = \theta_L - \theta_R$ and the respective equations for
$\rho_R$ and $\theta_R$ can be obtained by exchanging $L$ and $R$
and using $-\theta$ instead of $\theta$. Focusing now on the
steady solutions (satisfying
$\dot{\rho}_{L,R} = \dot{\theta}_{L,R} = 0$), we need to enforce
$\theta = 0 $ or $\pi$ for non-zero amplitudes. This leads us to
symmetric and antisymmetric (equal or opposite amplitudes) \emph{pairs} of solutions, namely for $\theta = 0$ we have the symmetric
(only positive ones among the) solutions $\rho_{L,R}^2 = \left(-s\eta_0 \pm \sqrt{\eta_0^2 - 4\delta\eta_4(\omega_0-\mu)}\right)/2\delta\eta_4$ with $\displaystyle{\mu < \omega_0 + \frac{\eta_0^2}{4\eta_4}}$, for $\delta = -1$ ( $\displaystyle{\mu > \omega_0 - \frac{\eta_0^2}{4\eta_4}}$ for $\delta=1$). Also,
for $\theta=\pi$, we have (only the positive amplitude
ones among) the antisymmetric solutions $\rho_{L,R}^2 = \left(-s\eta_0 \pm \sqrt{\eta_0^2 - 4\delta\eta_4(\omega_1-\mu)}\right)/2\delta\eta_4$ with $\displaystyle{\mu < \omega_1 + \frac{\eta_0^2}{4\eta_4}}$ for $\delta = -1$ (resp. $\displaystyle{\mu > \omega_1 - \frac{\eta_0^2}{4\eta_4}}$ for $\delta =1$). For the asymmetric solutions one has to solve the polynomial
\begin{eqnarray*}
\delta\eta_4\rho_{L,R}^6 + s\eta_0\rho_{L,R}^4 + (\Omega - \mu)\rho_{L,R}^2 + \frac{\omega^2}{s\eta_0 + \delta\eta_4N} = 0,
\end{eqnarray*}
which can more conveniently be written as a function of the norm
of the solutions (representing the atom number in BECs and the optical
intensity in optics). Thus, introducing $N$ ($N = \rho_L^2 + \rho_R^2$)
yields the quartic polynomial
\begin{eqnarray*}
\delta^3\eta_4^3N^4 + 3s\eta_4^2\eta_0N^3 + (3\delta\eta_4\eta_0^2 - \eta_4^2(\mu-\Omega))N^2 + (s^3\eta_0^3 - 2s\delta\eta_0\eta_4(\mu-\Omega))N - \delta\eta_4\omega^2 - \eta_0^2(\mu - \Omega) = 0.
\end{eqnarray*}

For the second case ($\sigma > \sigma_b$) the integrals $\eta_0$, $\eta_1$, $\eta_4$ are taken into account and the projection equations
onto the states $\phi_{L,R}$, read:
\begin{eqnarray*}
i\dot{c}_L = (\Omega - \mu)c_L - \omega c_R + s c_L(\eta_0|c_L|^2 + \eta_1|c_R|^2) + \delta\eta_4|c_L|^4c_L\\
i\dot{c}_R = (\Omega - \mu)c_R - \omega c_L + s c_R(\eta_0|c_R|^2 + \eta_1|c_L|^2) + \delta\eta_4|c_R|^4c_R.
\end{eqnarray*}
Here, the amplitude-phase decomposition yields
\begin{eqnarray*}
\left\{ \begin{array}{cc} \dot{\rho}_L = \omega\rho_R\sin\theta\hspace{4.6 cm}\\ \\
\dot{\theta}_L = \mu - \Omega + \omega\frac{\rho_R}{\rho_L}\cos\theta - s\eta_0\rho_L^2 - s\eta_1\rho_R^2 - \delta\eta_4\rho_L^4. \end{array} \right\}
\end{eqnarray*}
We can, once again, obtain the set of stationary solutions as follows.
When $\theta = 0$ (symmetric case) the solutions will be
(the positive amplitude ones among)
$\displaystyle{\rho_{L,R}^2 = \left(-s(\eta_0 + \eta_1) \pm\sqrt{(\eta_0 + \eta_1)^2 - 4\delta\eta_4(\omega_0-\mu)}\right)/2\delta\eta_4}$ for $\displaystyle{\mu < \omega_0 + \frac{(\eta_0 + \eta_1)^2}{4\eta_4}}$ for $\delta = -1$ ($\displaystyle{\mu > \omega_0 - \frac{(\eta_0 + \eta_1)^2}{4\eta_4}}$ for $\delta = 1$) and when $\theta = \pi$ (antisymmetric case) the solutions are
(the positive amplitude ones among) $\displaystyle{\rho_{L,R}^2 = \left(-s(\eta_0 + \eta_1) \pm\sqrt{(\eta_0 + \eta_1)^2 - 4\delta\eta_4(\omega_1-\mu)}\right)/2\delta\eta_4}$ and exist for $\displaystyle{\mu < \omega_1 + \frac{(\eta_0 + \eta_1)^2}{4\eta_4}}$ for $\delta = -1$ ($\displaystyle{\mu > \omega_1 - \frac{(\eta_0 + \eta_1)^2}{4\eta_4}}$ for $\delta = 1$). The asymmetric solutions now, directly in norm expression, will be given by the polynomial
\begin{eqnarray*}
\delta^3\eta_4^3N^4 + (3s\eta_4^2\eta - s\eta_4^2\eta_1)N^3 + (3\delta\eta_4\eta^2 - \eta_4^2(\mu-\Omega) - 2\delta\eta_4\eta\eta_1)N^2 +\\ + (s^3\eta^3 - 2s\delta\eta\eta_4(\mu-\Omega) -s\eta_1\eta_4^2)N - \delta\eta_4\omega^2 - \eta^2(\mu - \Omega) = 0
\end{eqnarray*}
with $\eta$ here standing for $\Delta\eta = \eta_0 - \eta_1$.

In the third case, when $\sigma > \sigma_c$, the effect of the
quintic terms is deemed to be negligible and the situation reverts
to the analysis of~\cite{chenyu} and is hence omitted here.



\subsection{The bifurcation analysis}

In order to derive a more convenient form of the system so that we can proceed to the analysis of the spontaneous symmetry breaking
(SSB) bifurcation, we introduce the population imbalance between the two
wells,
\begin{eqnarray}
z = (N_L - N_R)/N = (|c_L|^2 - |c_R|^2)/N,
\label{variable1}
\end{eqnarray}
where $N_{L,R} = |c_{L,R}|^2 = \rho_{L,R}^2$ and $N = N_L + N_R$. Together
with the relative phase between the two wells
$\theta = \theta_L - \theta_R$, this forms a set of
conjugate variables, in which we obtain the dynamical system :
\begin{eqnarray*}
\left\{ \begin{array}{cc} \dot{z} = 2\omega\sqrt{1 - z^2}\sin\theta\hspace{1.8 cm} \\ \dot{\theta} = \displaystyle{-\frac{2\omega z\cos\theta}{\sqrt{1 - z^2}} - s\eta Nz - \delta\eta_4N^2z.}\end{array} \right\}
\end{eqnarray*}
This can be written in the Hamiltonian form
\begin{eqnarray}
\left\{\begin{array}{cc} \dot{z} = \displaystyle{-\frac{\partial {\cal H}}{\partial\theta}}\\
\\
\dot{\theta} = \displaystyle{\frac{\partial {\cal H}}{\partial z}} \end{array}
\right\}
\label{system}
\end{eqnarray}
with the Hamiltonian function
\begin{eqnarray*}
{\cal H} = 2\omega\sqrt{1 - z^2}\cos\theta - \frac{1}{2}s\eta Nz^2 - \frac{1}{2}\delta\eta_4N^2z^2.
\end{eqnarray*}
Note that $\eta$ stands either for $\eta_0$ ($\sigma<\sigma_b$) or for
$\Delta\eta = \eta_0 - \eta_1$ ($\sigma_b<\sigma<\sigma_c$). The system
possesses the stationary solutions (critical points) $(z_1,\theta_1)$ and
$(z_2,\theta_2)$ with $z_1 = z_2 = 0$, $\theta_1 = 0$, $\theta_2 = \pi$ that
correspond to the symmetric and antisymmetric solutions, identified
above. Furthermore, the stationary solutions representing the asymmetric branches are given by:
\begin{eqnarray*}
z^2 = 1 - \frac{4\omega^2}{(s\eta N + \delta\eta_4N^2)^2}, \ \ \theta =
0,\ \pi.
\end{eqnarray*}
These branches emerge and merge as bifurcations from and to the symmetric or antisymmetric solutions and they exist for those values of $N$ for which $z^2 \geq 0$. Taking $z = 0$, we get that
\begin{eqnarray}
N = (-s\eta \pm \sqrt{\eta^2 + 8\delta
\eta_4\omega})/2\delta\eta_4, \ \ N = (-s\eta \pm \sqrt{\eta^2 - 8\delta\eta_4\omega})/2\delta\eta_4.
\label{ns}
\end{eqnarray}
By substituting $(s,\delta) = (1,-1)$ or $(-1,1)$ we get the same four possible expressions for $N$ as a function of $\sigma$ that are displayed in fig.$4$ and we denote them with $N_0^{cr}$, $N_1^{cr}$, $N_2^{cr}$ and $N_3^{cr}$
(the subscripts $0$ and $2$ correspond to the (-) signs in the
left and right expressions of Eq.~(\ref{ns}), respectively, while the
subscripts $1$ and $3$ to the (+) signs). One can then see that when $(s,\delta) = (1,-1)$ and demanding that $z^2 > 0$, one gets that $N$ should either lie in the area outside the curves $N_0^{cr}$ and $N_1^{cr}$ or in the area inside the curves $N_2^{cr}$ and $N_3^{cr}$.
In the case of $s=-1$ and $\delta=1$,
the role of the symmetric and anti-symmetric branches
gets exchanged in as far as the bifurcation of the asymmetric branch is
concerned (see also below).


\begin{figure}[th]
\begin{center}
\includegraphics[width = 8cm]{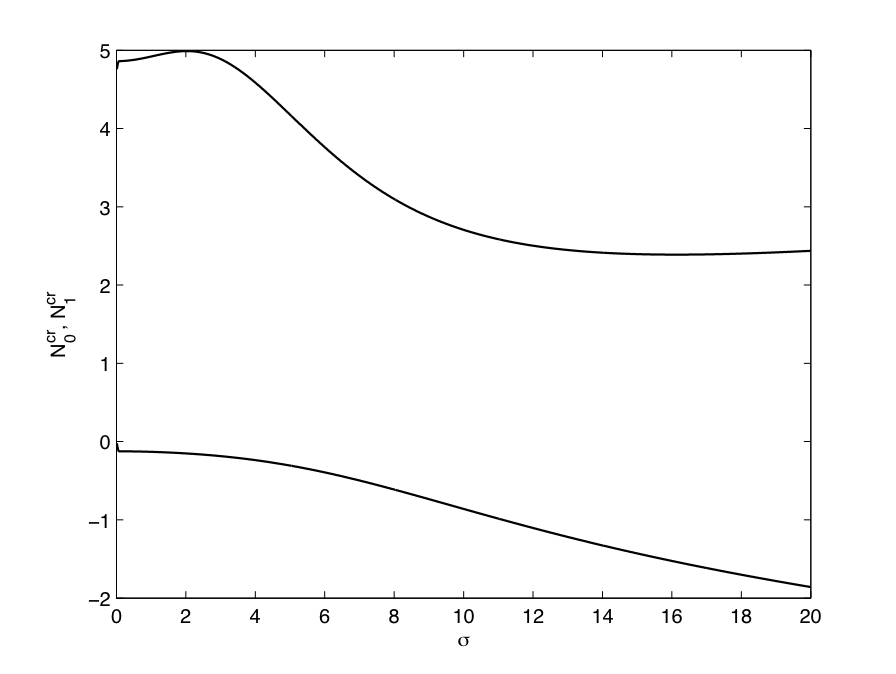}
\includegraphics[width = 8cm]{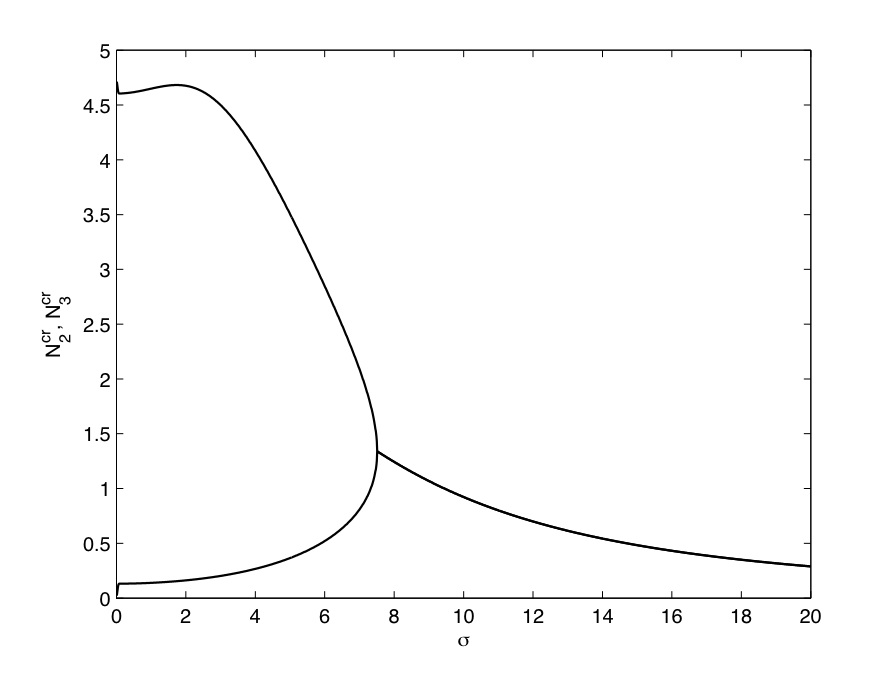}
\end{center}
\caption{ The critical values $N_0^{cr}$, $N_1^{cr}$ (left panel) and $N_2^{cr}$, $N_3^{cr}$ (right panel) whenever $(s,\delta) = (1,-1)$ show when the
bifurcations appear. More specificaly, the left panel corresponds to the bifurcations that occur on the symmetric branch and the right panel for those that occur on the antisymmetric one.
}
\label{crit_points}
\end{figure}

Importantly, it can be observed in Fig.~\ref{crit_points} that
$N_0^{cr}$ is always negative, hence it is omitted for the principal
case considered herein, namely $s=1$ and $\delta=-1$. On the one hand, the {\it critical} conclusion of
our analysis is that for $\sigma < 7.52$, the system is predicted to have
for the anti-symmetric branch
both a symmetry breaking bifurcation (at $N= N_2^{cr}$) and
a symmetry restoring one that eliminates the asymmetric branch
(at $N=N_3^{cr}$). On the other hand, the right panel suggests that $N_2^{cr}$, $N_3^{cr}$ coincide $\sigma \geq  7.52$, beyond which there is only a single (symmetry breaking)
bifurcation. However, as will be discussed below, for large interaction range $\sigma$ this prediction seems to have some discrepancy from what actually happens as we will see that in fact, we observe a symmetry restoring
bifurcation
while we do not observe a bifurcation at all in the symmetric branch. To the best of our knowledge, this is the first example of an analytical prediction of the existence of a symmetry restoring bifurcation, a feature that is unique to the
analysis of the normal form of the bifurcation for the cubic-quintic
case (and cannot be predicted e.g. in the purely cubic case two-mode
analysis of~\cite{chenyu}).
The new critical points appear or disappear as a pitchfork bifurcation that
emerges from the antisymmetric solutions for
$\theta = \pi$ respectively. From the symmetric solution, in this
case of $s=1$ and $\delta=-1$, only a single bifurcation arises
at $N=N_1^{cr}$.

For the opposite case (to the one principally considered herein)
of $s=-1$ and $\delta=1$, i.e., for a focusing
cubic nonlinearity, the bifurcations emerge from the symmetric branch,
while for $s=1$, i.e., for a defocusing cubic term, then the relevant
symmetry breakings arose from the anti-symmetric branch.
Thus, in this case, we expect an asymmetric branch to bifurcate
and break the symmetry at $N= N_2^{cr}$, while it returns to the parent
symmetric branch restoring the symmetry at $N= N_3^{cr}$. On the other hand,
for the anti-symmetric waveform with a focusing cubic nonlinearity,
only a single bifurcation arises at $N= N_1^{cr}$. We provide further
details of each of these bifurcations and their comparison with the
full numerics of the underlying NLS model in the next section.

\begin{figure}[th]
\begin{center}
\includegraphics[width = 7 cm]{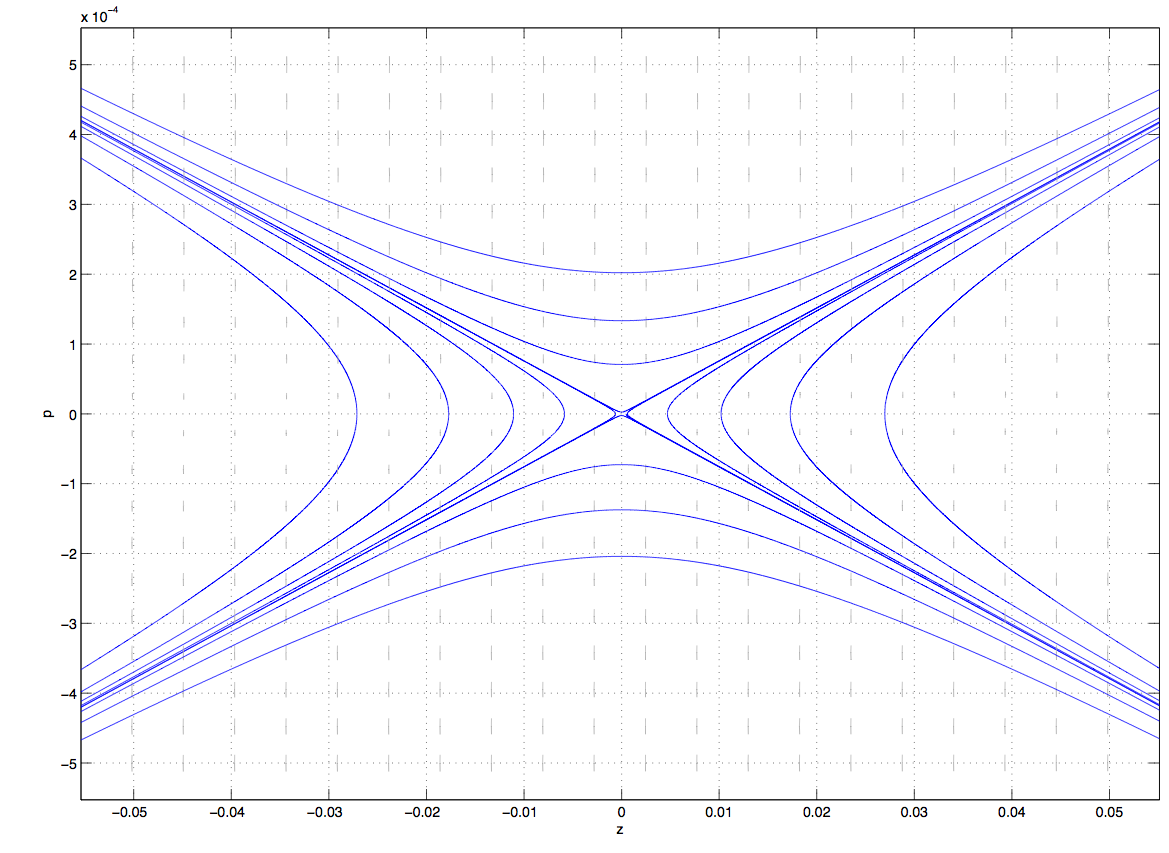}
\includegraphics[width = 7 cm]{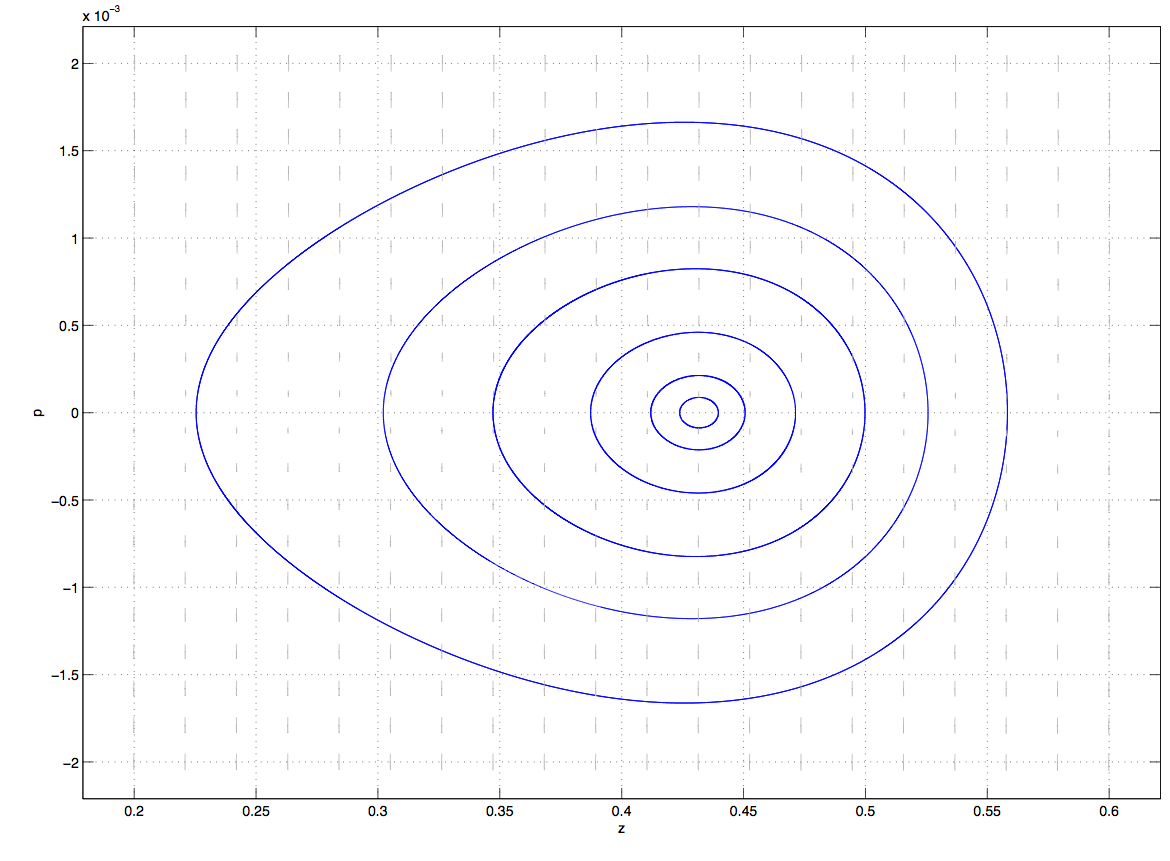}
\includegraphics[width = 10 cm]{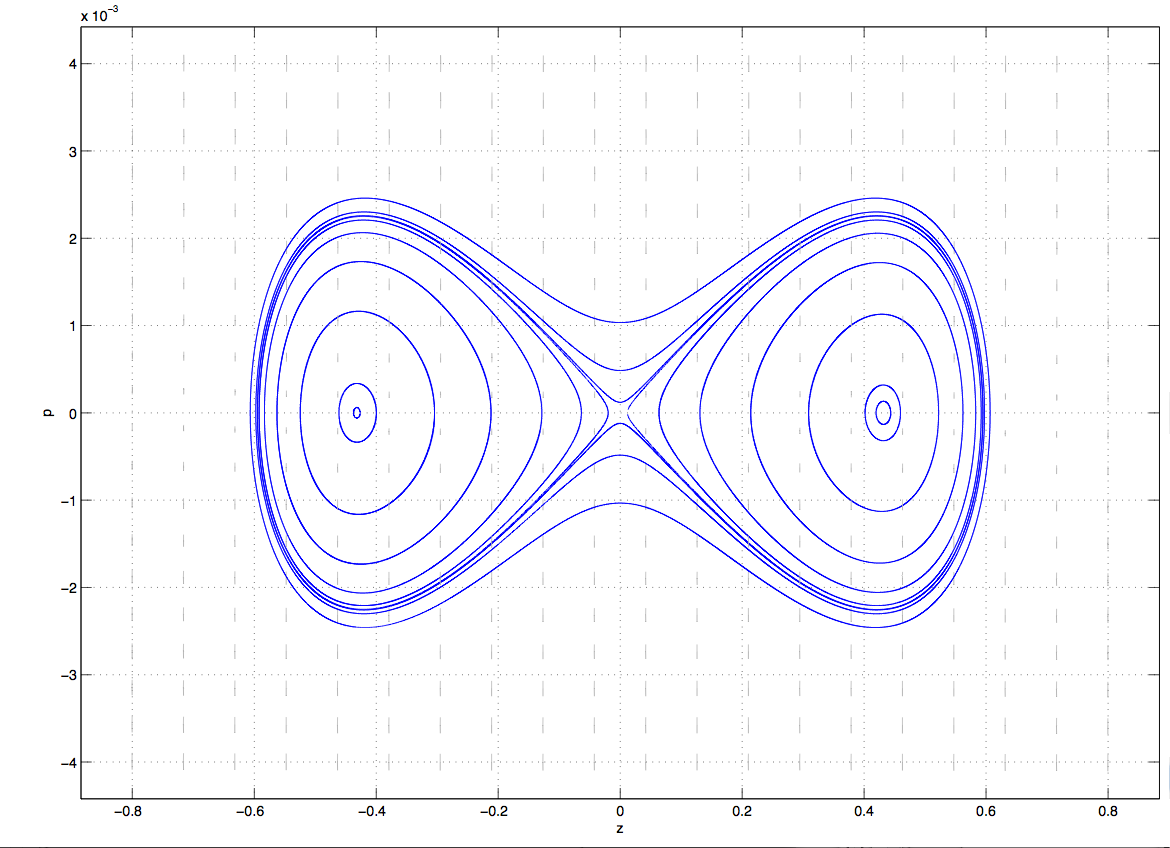}
\end{center}
\caption{
Top panels: The phase space diagrams of the Hamiltonian system when $s = 1$ and
$\delta = -1$ with the Gaussian kernel, for $\sigma = 1$, $N = 5$, and
with $N_1^{cr} = 4.9862$ (after the new fixed points are created at
$(\pm 0.4318,0)$). The left panel displays the region of phase space
near the symmetric solution
$(0,0)$ (saddle) and the right panel the one near one of the
asymmetric fixed points $(0.4318,0)$ (center).
The bottom panel shows the full
phase space diagram of the system for $N=5$. }
\label{phase_plane}
\end{figure}

From the system of Eqs.~(\ref{system}), one can reduce the
dynamical evolution to a single second-order ODE:
\begin{eqnarray*}
\ddot{z} = -4\omega^2z - (s\eta Nz + \delta\eta_4N^2z)\sqrt{4\omega^2 - 4\omega^2z^2 - \dot{z}^2}
\end{eqnarray*}
which can also be rewritten in the ``position-momentum'' variables as:
\begin{eqnarray}
\left\{ \begin{array}{cc} \dot{z} = p,\hspace{6 cm}\\ \dot{p} = -4\omega^2z - (s\eta Nz + \delta\eta_4N^2z)\sqrt{4\omega^2 - 4\omega^2z^2 - p^2}.\end{array}
\right\}
\end{eqnarray}
This renders the system amenable to the phase plane representation
of the form shown in Fig.~\ref{phase_plane}.
Here we observe that there is a stationary solution $\dot{z} = \dot{p} = 0$
which is a fixed point of the center type. However, for the cases
when $(s,\delta) = (1,-1)$, for $N$ crossing the critical point
$N_1^{cr}$ in the case of the symmetric branch
and for $N \in [N_2^{cr},N_3^{cr}]$ in the case of the anti-symmetric branch,
there appear two more critical points at $p = 0$ and $\displaystyle{z = \pm \sqrt{1 - \frac{4\omega^2}{(s\eta N + \delta\eta_4N^2)^2}}}$, representing the asymmetric solutions. The point $(0,0)$ is a fixed point of center type before the bifurcation occurs, but past the relevant critical
number of atoms (or optical intensity), it becomes a saddle as the two new
(asymmetric) fixed points that appear are of center type.
Fig.~\ref{phase_plane} shows the phase space of the full system,
as well as the vicinity of the critical points for the Gaussian kernel
with $\sigma=1$, $N_1^{cr} = 4.9862$ and  $N = 5$. 

\begin{figure}[th]
\begin{center}
\includegraphics[width = 18cm]{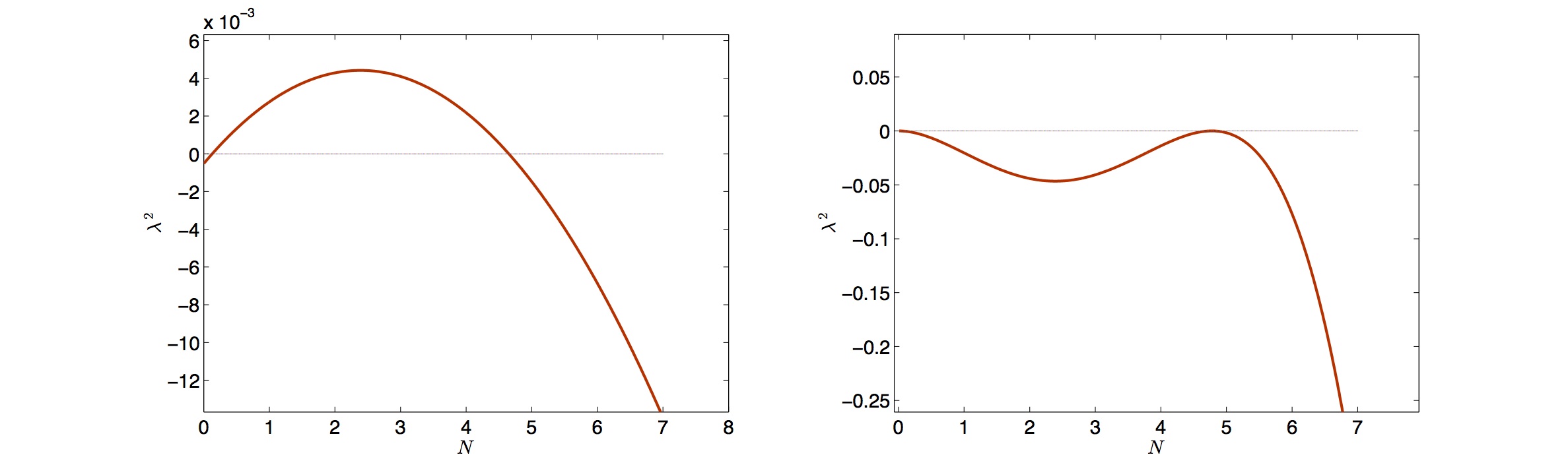}
\end{center}
\caption{ Graphs of the squared linearization eigenvalue
$\lambda^2$ vs. $N$ for the antisymmetric
  (left panel) and the asymmetric (right panel) stationary solutions
  when $\sigma = 0.1$. }
\label{eigen}
\end{figure}

It is worth mentioning at this point that there are no further changes in the
stability of the critical points (and thus in the corresponding stationary
solutions) for other values of $N$ except for those reported above. 
For the sake of simplicity we illustrate this below
for the antisymmetric solution bifurcation as a preamble towards the
corresponding
numerical results of the next section. The antisymmetric solution
corresponds the critical point $(z,p) = (0,\pi)$ where the Jacobian of the
linearized version of (\ref{system}) is  
\begin{eqnarray*}
J(z,\theta)|_{(0,\pi)} = \left(\begin{array}{cc}0 & -2\omega \\2\omega
    -(s\eta N + \delta\eta_4 N^2) & 0\end{array}\right)
\end{eqnarray*}
and its eigenvalues satisfy 
\begin{eqnarray*}
\lambda^2 + 4\omega^2 - 2\omega(s\eta N + \delta\eta_4 N^2) = 0.
\end{eqnarray*}
For the case where $\sigma = 0.1$ the graph of $\lambda^2$ versus $N$
(illustrated in the left panel of Fig.~\ref{eigen}) shows clearly that the two 
initially (i.e., close to the linear limit) purely imaginary 
eigenvalues
become real at $N = 0.14$, so that the center type equilibrium becomes a
saddle until $N = 4.63$ where it turns back to its initial state,
restoring the stability on the antisymmetric branch (symmetry-restoring
bifurcation) with no other
changes in between (or after that). 
For the asymmetric solution that corresponds to the
point $(z_0,\theta)$, where $z_0^2 = \displaystyle{1 -
  \frac{4\omega^2}{(s\eta N+\delta\eta_4 N^2)^2}}$ and $\theta = \pi$,
the Jacobian becomes 
\begin{eqnarray*}
J(z,\theta)|_{(z_0,\pi)} = \left(\begin{array}{cc}0 & \displaystyle{-\frac{4\omega^2}{s\eta N +
      \delta\eta_4 N^2}}\\\displaystyle{\frac{(s\eta N +\delta\eta_4 N^2)^3
    }{4\omega^2}\left(1 - \frac{4\omega^2}{s\eta N + \delta\eta_4 N^2}\right) }& 0\end{array}\right)
\end{eqnarray*}
and for its eigenvalues we obtain
\begin{eqnarray*}
\lambda^2 + \left(s\eta N + \delta\eta_4 N^2\right)\left(1 -
  \displaystyle{\frac{4\omega^2}{s\eta N + \delta\eta_4 N^2}}\right) = 0.
\end{eqnarray*}
Again for $\sigma = 0.1$ the graph of $\lambda^2$ versus $N$
(illustrated in the right panel of Fig.~\ref{eigen}) 
shows that the eigenvalues are always purely
imaginary which corresponds to an equilibrium of the center type. One can
observe here that the values of $N$ where the eigenvalues of
the asymmetric branch ``touch" the
$x$-axis ($N_1 = 0.03$ and $N_2 = 4.75$) coincide with the values
where the bifurcation occurs (Fig.~\ref{crit_points}-right panel) therefore the critical
points $z$ cease to exist before $N_1$ and after $N_2$. For the values of
$N$ within this interval, no further change of stability is
observed. As it is made clear in the next section, these stability
results are in excellent agreement with the corresponding numerical
ones. Additionally, it will be come transparent therein that additional
turning points in the $N$ vs. $\mu$ bifurcation diagram do {\it not}
correspond to any instabilities in complete agreement with the recent
analysis of~\cite{jyang}.

\section{Numerical Approach}


\subsection{Stationary solutions}

We now turn to the examination of our analysis against
the results of numerical bifurcation analysis (and in the next
subsection also compare them to direct numerical simulations).
We focus here on the case where $s=1$, $\delta = -1$, as we are
especially interested in the case of \emph{competing} interactions;
we will briefly also touch upon the case of $s=-1$ and $\delta=1$.
In our numerical computations, the stationary solutions are obtained by
using a fixed-point Newton-Raphson iteration for a finite difference
decomposition of the relevant boundary value problem, with a choice
of the grid spacing of $\Delta x = 0.1$ and employing a parametric
(and wherever needed a pseudo-arclength) continuation of the solutions with
respect to the chemical potential parameter $\mu$ (in optics this is the
so-called propagation constant). The linear stability is analyzed by
considering the standard linearization around the
stationary solutions $\psi_0$ in the form
\begin{eqnarray*}
\psi(x,t) = \psi_0 + \epsilon(a(x)e^{\lambda t} + b^*(x)e^{\lambda^*t}).
\end{eqnarray*}
This yields the eigenvalue problem
\begin{eqnarray*}
\left(\begin{array}{cc}L_1 & L_2\\ - L_2^* & -L_1^*\end{array}\right)\left(\begin{array}{c}a\\b\end{array}\right) = i\lambda\left(\begin{array}{c}a\\b\end{array}\right),
\end{eqnarray*}
where the operators are defined as
\begin{eqnarray*}
L_1\phi = \left[-\frac{1}{2}\partial_x^2 + V - \mu +s\int_{-\infty}^{+\infty}K(x-x')|\psi_0(x')|^2dx' + \delta\int_{-\infty}^{+\infty}K(x-x')|\psi_0(x')|^4dx'\right]\phi + \\ + s\int_{-\infty}^{+\infty}K(x-x')\psi_0(x)\psi_0^*(x')\phi(x')dx' + 2\delta\int_{-\infty}^{+\infty}K(x-x')\psi_0(x)\psi_0(x'){\psi_0^*}^2(x')\phi(x')dx'
\end{eqnarray*}
and
\begin{eqnarray*}
L_2\phi = s\int_{-\infty}^{+\infty}K(x-x')\psi_0(x')\psi_0(x)\phi(x')dx + 2\delta\int_{-\infty}^{+\infty}K(x-x')\psi_0(x)\psi_0^*(x')\psi_0^2(x')\phi(x')dx'
\end{eqnarray*}
for any real function $\phi$. Instability is guaranteed by the existence
of any eigenvalues $\lambda$ of the linearized operator
with $\Re (\lambda) \neq 0$ in the sense that perturbations along the
corresponding eigendirections will deviate exponentially from the
corresponding fixed point.
Recall that this is also the case for all eigenvalues of our Hamiltonian system, since when $\lambda$ is an eigenvalue, so are $-\lambda$, $\lambda^{\star}$
and $-\lambda^{\star}$. In the case where all eigenvalues are found to
be purely imaginary, then
the solution is found to be marginally
stable.

\begin{figure}[th]\label{stat}
\begin{center}
\includegraphics[width = 7cm]{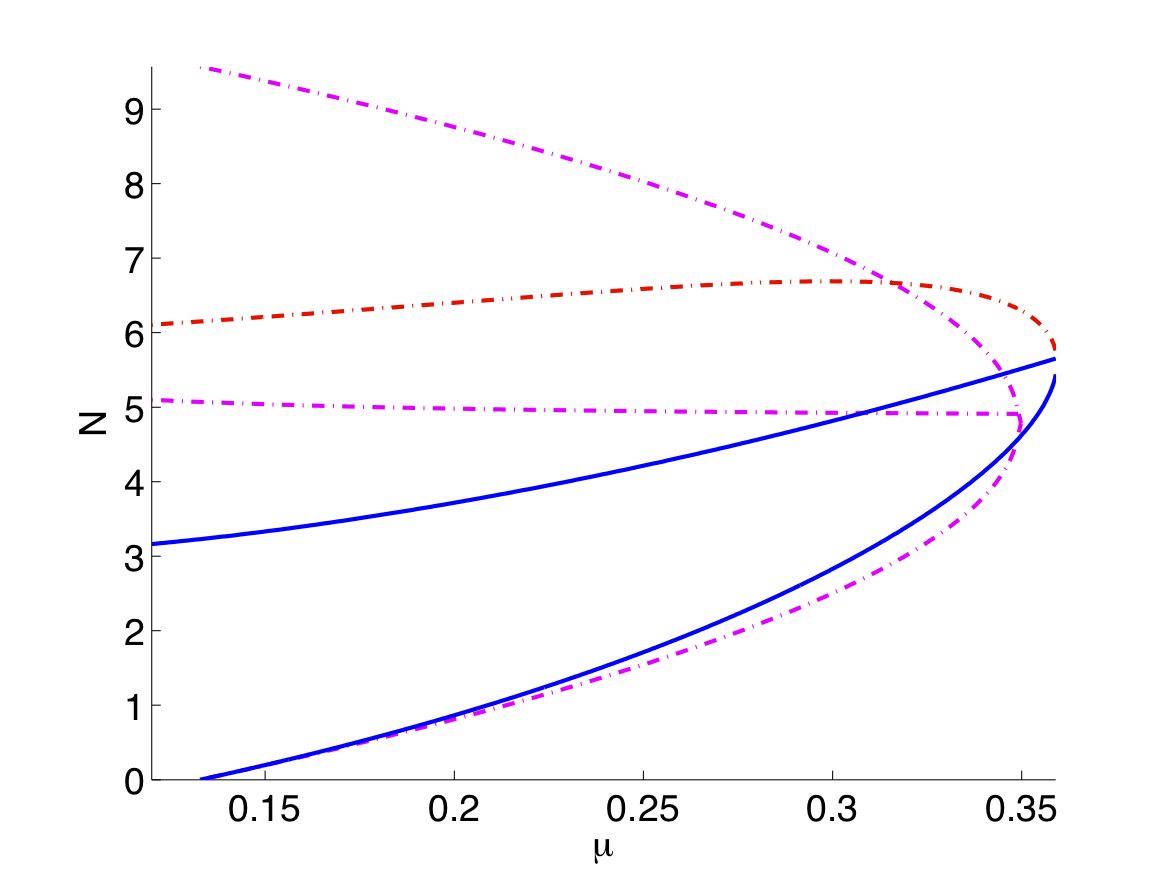}
\includegraphics[width = 7cm]{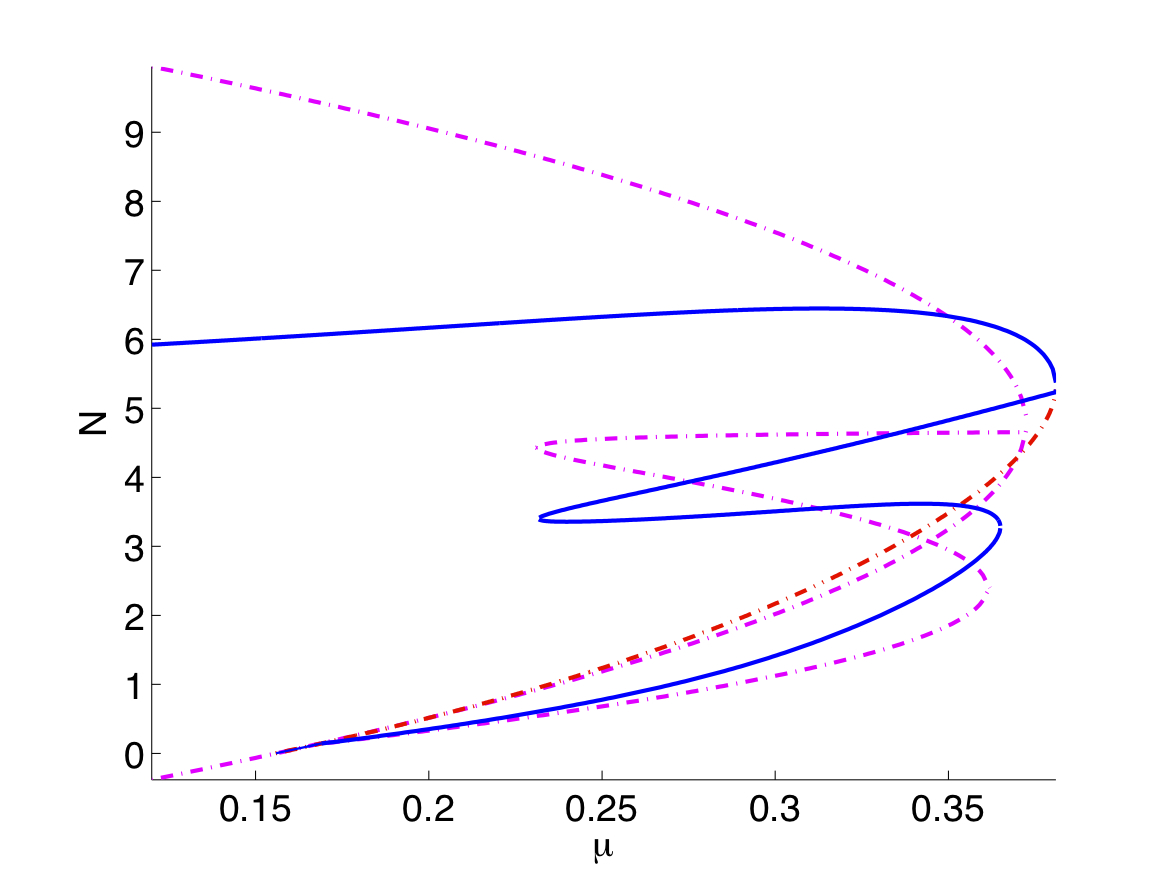}\\
\includegraphics[width = 7cm]{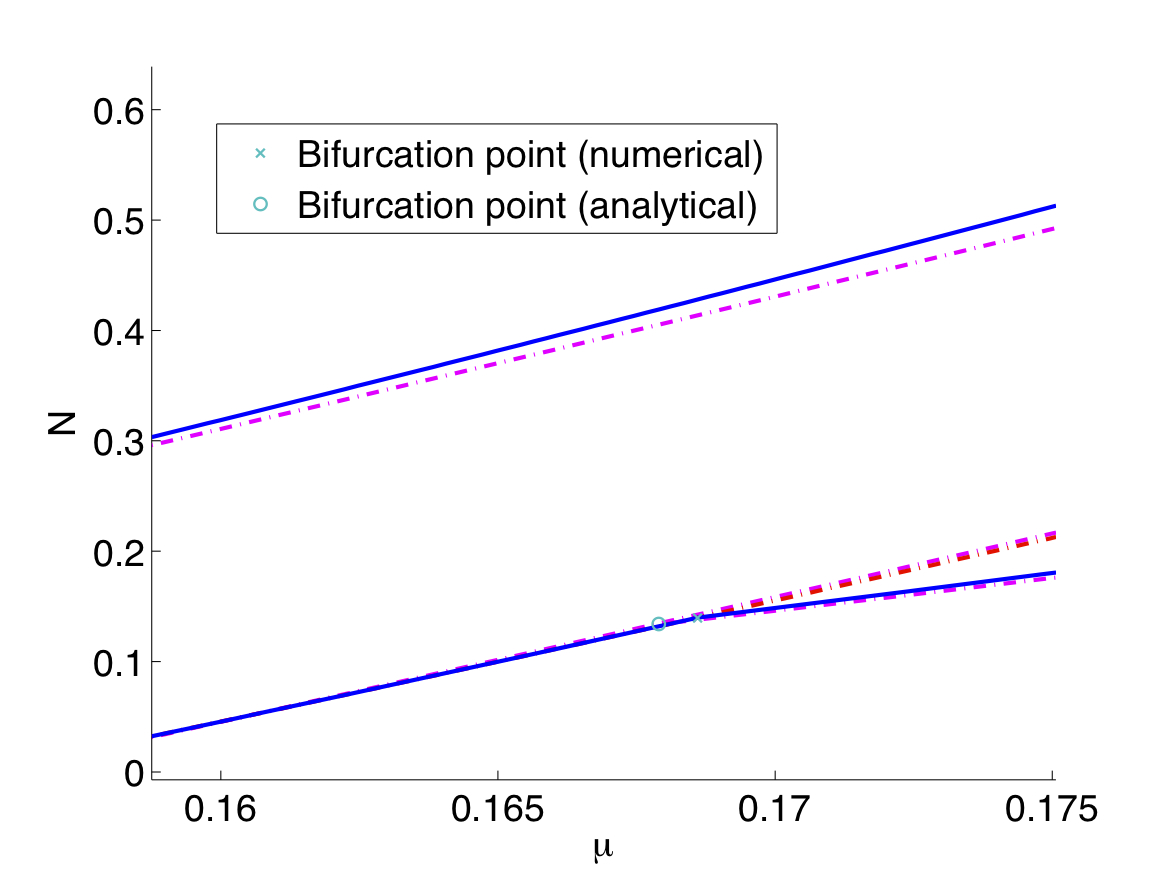}
\end{center}
\caption{The stationary solution branches
for the case $s=1$, $\delta = -1$ when the interaction range is $\sigma = 0.1$
expressed in terms of the normalized $N$ as a function of $\mu$.
The analytical predictions are denoted with the purple dash--dotted line while
the numerically determined solutions are denoted with the solid line that is
blue when it is stable and red otherwise. The top left panel shows
the symmetric solutions, while the top right presents the antisymmetric ones,
both including the asymmetric bifurcations that emerge from them.
The bottom panel presents a detail of the symmetry-breaking effect,
showcasing the quality of its approximation by the two-mode expansion.}
\label{stat1}
\end{figure}

\begin{figure}
\begin{center}
\includegraphics[width = 7cm]{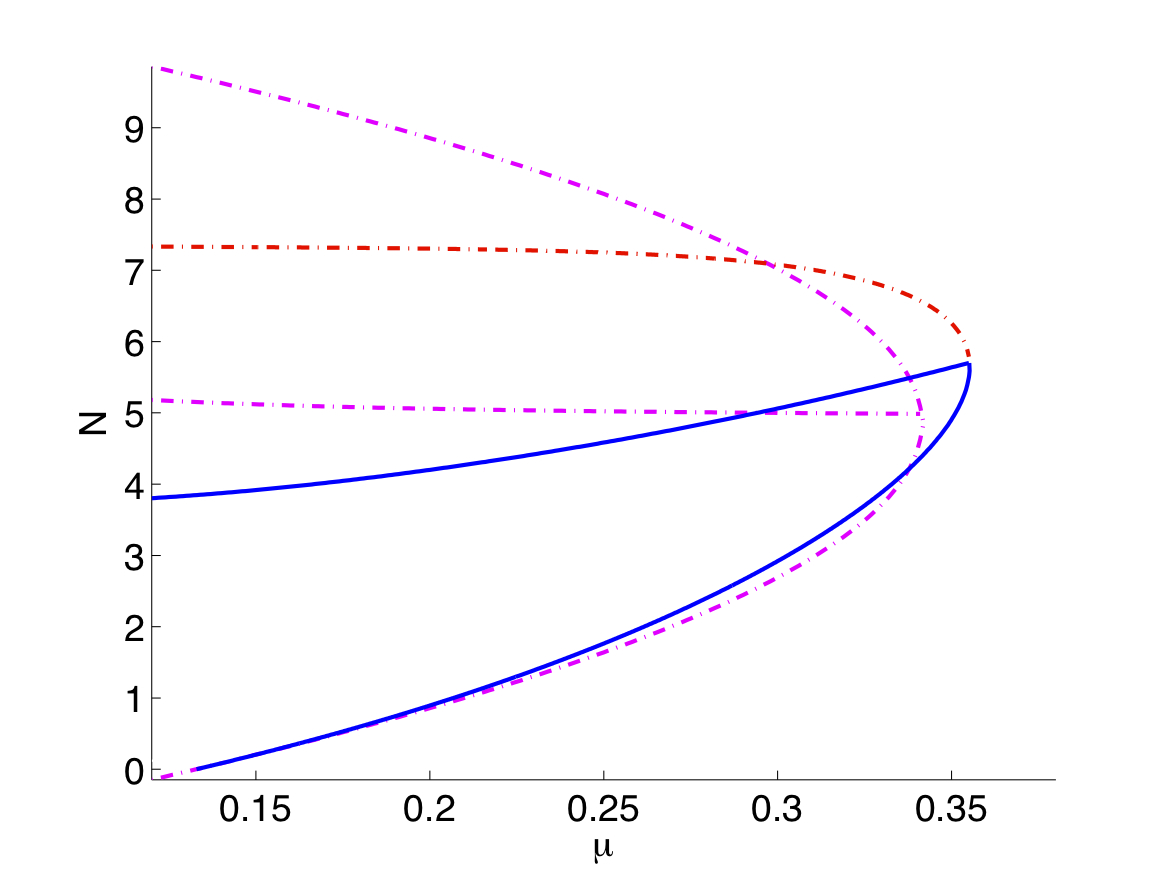}
\includegraphics[width = 7cm]{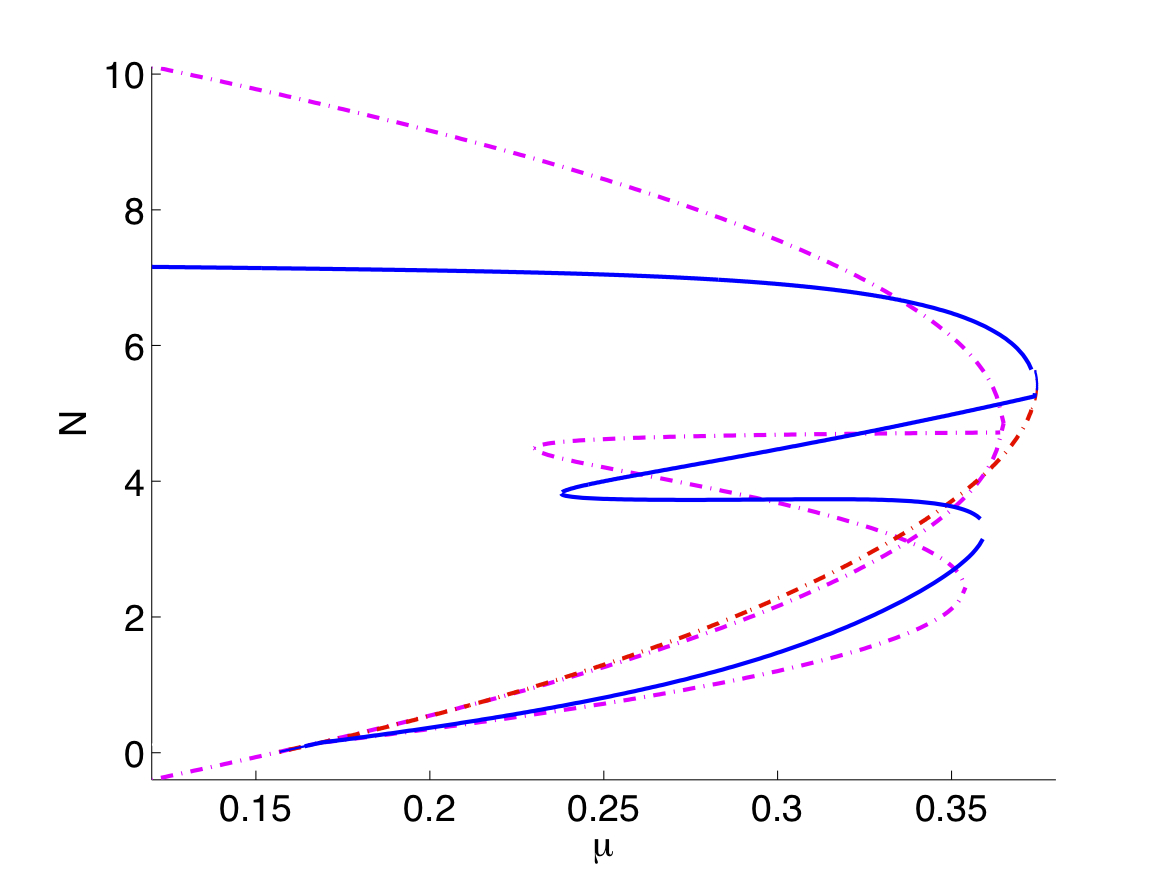}\\
\includegraphics[width = 7cm]{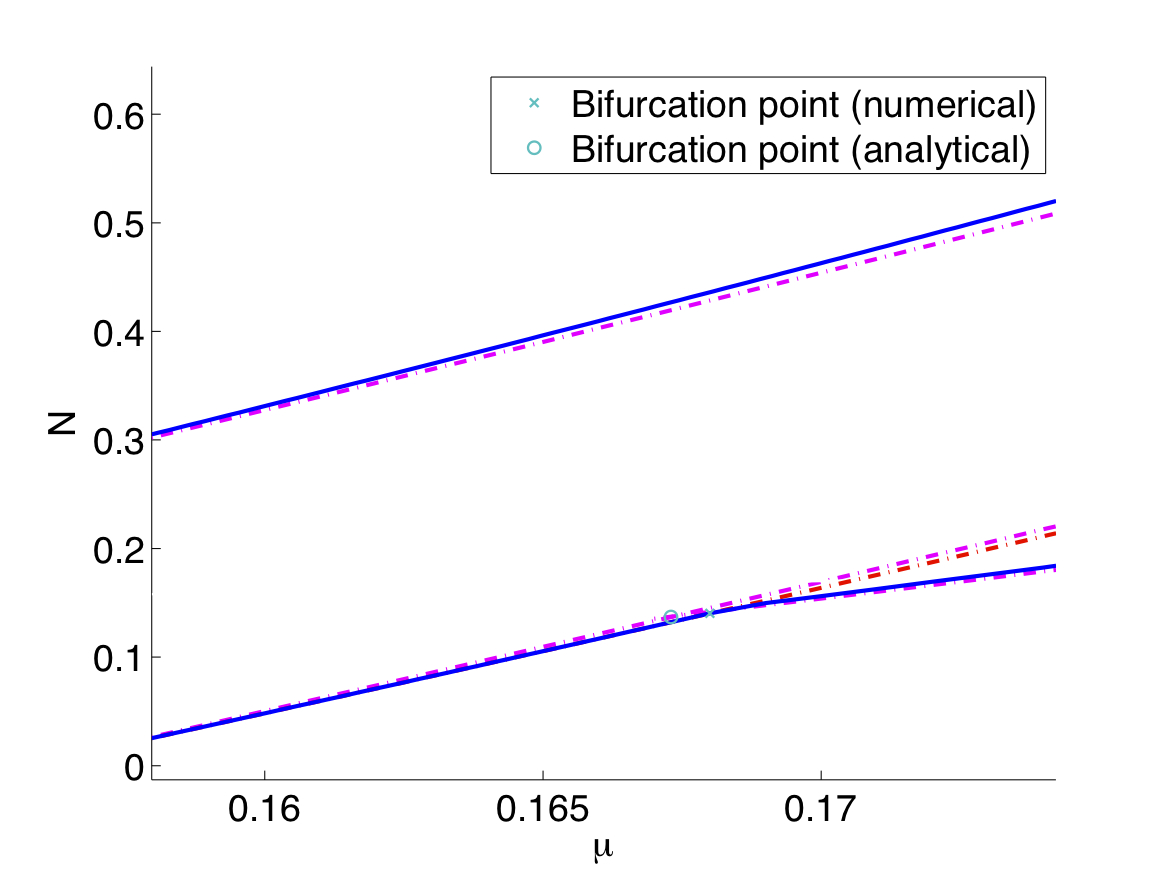}
\end{center}
\caption{This figure shows the same features as the previous
one for the symmetric branch
(top left panel), the anti-symmetric branch (top right panel)
and a zoom-in of the symmetry breaking (bottom panel). However,
the interaction range here is an order of magnitude larger,
namely $\sigma=1$.}
\label{stat2}
\end{figure}

\begin{figure}
\begin{center}
\includegraphics[width = 7cm]{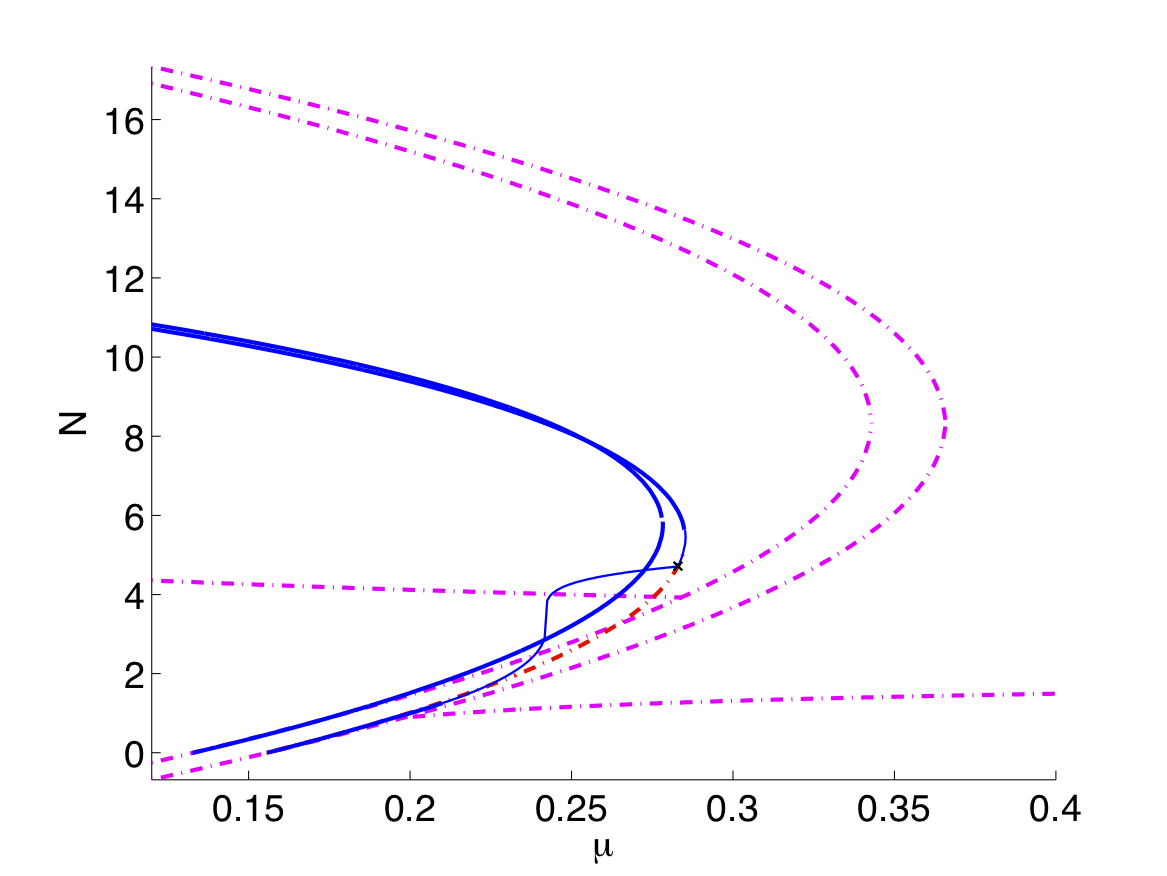}
\includegraphics[width = 7cm]{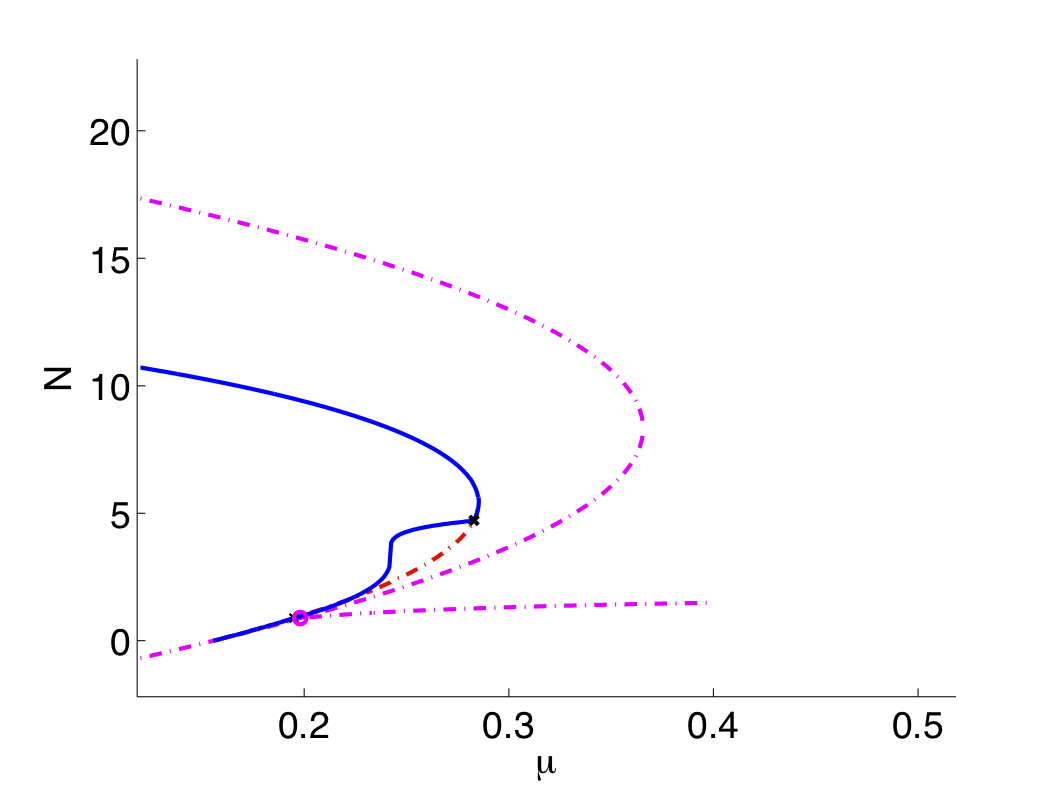}\\
\includegraphics[width = 7.5cm]{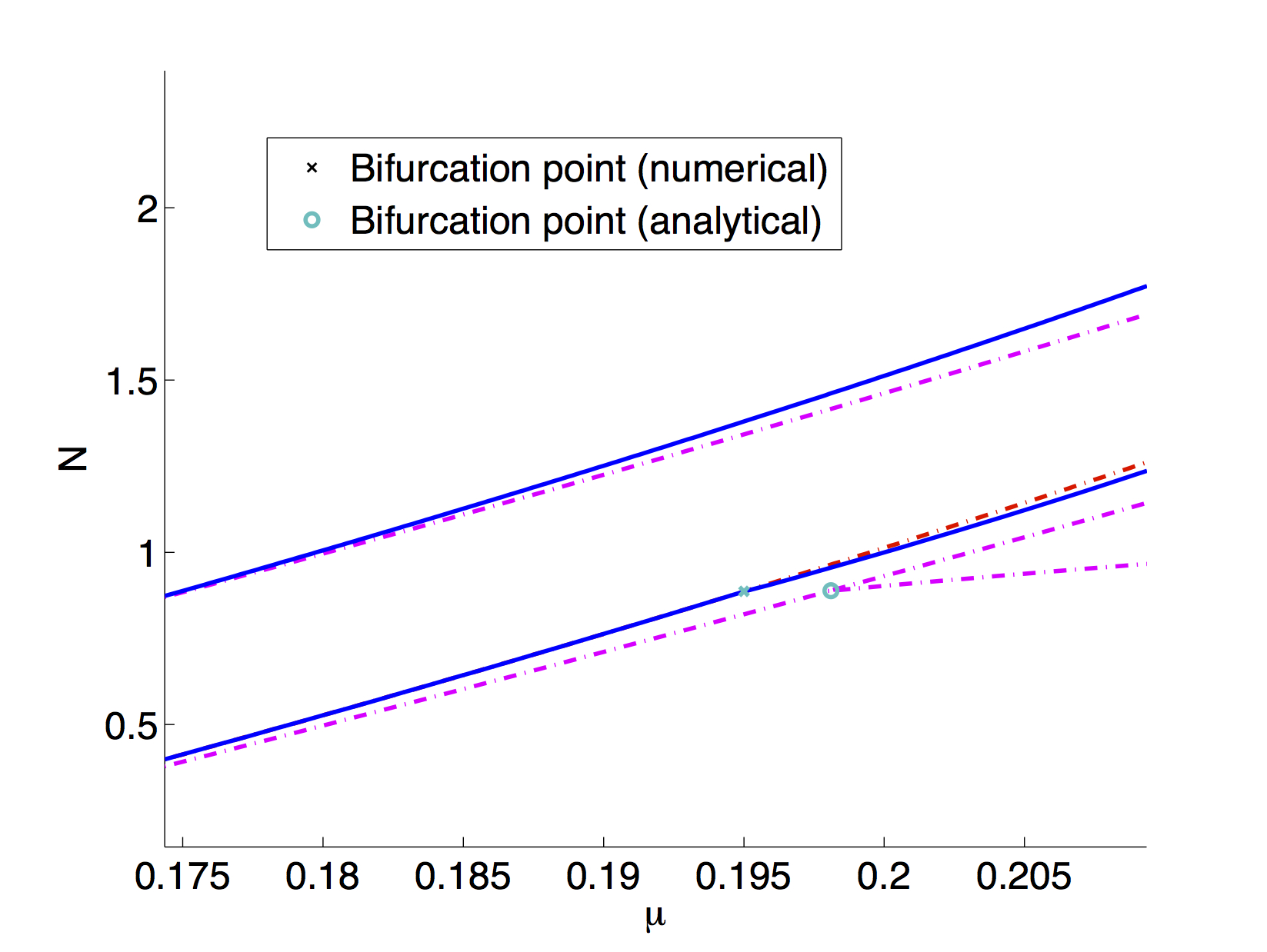}
\end{center}
\caption{Same as the previous two figures, but now for large
nonlocality interaction range in the case of $\sigma=8$.}
\label{stat3}
\end{figure}

In our specific case of competing interactions, we comment on the following.
The positive value ($s = 1$) denotes the repulsive behavior of the cubic
nonlocal term while the negative one $\delta = -1$ leads to
attractive behavior of the quintic nonlocal nonlinearity. As we examine
the bifurcation problem of nonlinear states from the corresponding
linear eigenstates, we expect that for lower values of $N$ (i.e., weaker
nonlinearities), the former repulsive term should be dominant, while
for larger values of $N$ (i.e., stronger nonlinearities), it is
anticipated that the latter attractive term will take over. This
is accurately reflected in the numerical bifurcation diagrams that
we now show in Figs.~\ref{stat1}-\ref{stat3},
for three (distinct by roughly an order of magnitude in each
case) values of the range $\sigma$. The first value of $\sigma=0.1$
in Fig.~\ref{stat1}
is supposed to reflect the local case (since the range of interaction
is much smaller than any other intrinsic length scale in the system).
Here the agreement with the two-mode approximation is very good quantitatively
for low $N$ and very good qualitatively (and even good quantitatively for
some features such as chemical potentials of critical points) for
large $N$. The quality of these types of agreements is found to be
preserved for an intermediate interaction range of $\sigma=1$
in Fig.~\ref{stat2}. However, when the interaction range becomes
sufficiently large that it competes (or overcomes) the length scale
of the potential wells, then fundamental disparities are expected
to be found and that is the very conclusion of Fig.~\ref{stat3} for
$\sigma=8$.

In the first case where $\sigma = 0.1$, the symmetric and antisymmetric
branches of nonlinear states emanate from $\mu = 0.1328$ and $0.1557$,
as expected, respectively ($\omega_0$ and $\omega_1$), both of them being
dynamically stable, for sufficiently small values of $N$. The rightward
bending of the branches for small $N$ confirms the dominance of the
self-repulsive part of the (cubic) interactions for small $N$, as
indicated above.
The antisymmetric branch (top right panel of Fig.~\ref{stat1} and
see also the zoom of the bottom panel of the figure)
is destabilized and the theoretically predicted asymmetric branch emerges.
The numerical value of the chemical potential for the bifurcation point is
found to be $\mu = 0.1686$, whereas the corresponding analytical one is
$\mu = 0.1679$, confirming the quantitative nature of the agreement with
the two-mode approximation. For larger $N$, we observe that the asymmetric
solution
has two apparent turning points (where the sign of $dN/d\mu$ changes, but
in fact its stability does not change - which agrees with the
theoretical result presented in the previous section), before it reaches the
anti-symmetric branch at the numerically computed value $\mu = 0.381$ where
we observe the symmetry restoring effect, which, in fact, re-stabilizes
the anti-symmetric branch.
In our theoretical analysis, we observe the same qualitative behavior and
the symmetry restoring occurs at $\mu = 0.3723$, in reasonable
agreement with the full numerical results. Two additional observations
should be made here. On the one hand, since the symmetry restoring
occurs at much larger values of $N$, the relevant agreement is expected
to be less adequate quantitatively than for the symmetry breaking occurring
at lower $N$. This is because a two-mode expansion is less appropriate
of a reduction at such higher nonlinearities. On the other hand,
it can indeed be observed that while the overall trend of the two curves
is the same (and even critical/turning points in terms of their chemical
potential are rather accurately captured), this agreement is not
adequate quantitatively e.g. for critical values of $N$ (or for detailed
quantitative matching of the curves for large $N$).
For the symmetric solution of the top left panel of Fig.~\ref{stat1},
we can observe that it is increasing monotonically until $\mu = 0.359$ where
it sustains a pitchfork bifurcation leading to the emergence
of an asymmetric branch and also a subsequent turning point.
The symmetric branch becomes unstable thereafter and the asymmetric
emerging state is the stable daughter branch. Notice that the theoretical
analysis is once again quantitatively accurate for small $N$ and the agreement
becomes more qualitative for higher $N$'s. The critical point for the emergence
of the asymmetric branch is predicted for $\mu=0.3492$ in reasonable
agreement with the full numerical result.

For the case of $\sigma = 1$ the effects are similar to those in the previous case. The symmetry breaking of the antisymmetric branch (top right, as well
as zoom in of the bottom panel of Fig.~\ref{stat2}) occurs now at
$\mu = 0.168$ according to the numerical results and at $\mu = 0.1673$
in the two-mode approximation, again attesting to its validity for
small $N$. After following a similar trajectory with the case $\sigma = 0.1$, the asymmetric solution merges back to the antisymmetric one at $\mu = 0.374$
(numerical value) or at $\mu = 0.364$ (analytical value) with the
antisymmetric branch again regaining its stability past the
symmetry restoring bifurcation. The symmetric solution (top left
panel of Fig.~\ref{stat2}) again increases
monotonically until it sustains a symmetry breaking bifurcation of its
own at $\mu = 0.355$. The two-mode approximation predicts this bifurcation
to arise at $\mu = 0.342$.

Next, in Fig.~\ref{stat3}, we increase the interaction range, roughly,
another order of magnitude by setting $\sigma = 8$. Here, as may be
intuitively expected given that the interaction range is wider
than the wells of the potential, the results are quite different.
For small values of $\mu$ (and thus atom number $N$ or optical power)
we have a quite satisfactory agreement (even quantititative) with
the two mode approximation, as may be expected. As a demonstration
of that, we note that the symmetry breaking of the antisymmetric branch
occurs in our analysis at
$\mu = 0.1981$, while numerically it is found to take place
at $\mu = 0.195$. On the other hand, due to the predicted earlier
collision of the critical points $N_2^{cr}$ and $N_3^{cr}$, there
is no symmetry restoring taking place in our normal form reduction.
Nevertheless, we observe that such a restoring, in fact, still takes
place in the full numerical bifurcation diagram.
Furthermore, in this case, we have not
been able to detect a symmetry-breaking bifurcation in the case
of the symmetric branch, even though such a bifurcation is predicted
within the reduction.
This illustrates that
for such large values of $\sigma$, even the qualitative agreement
previously associated with the large $N$ case dynamics should not
be expected to be present.

\begin{figure}
\begin{center}
\includegraphics[width = 7cm]{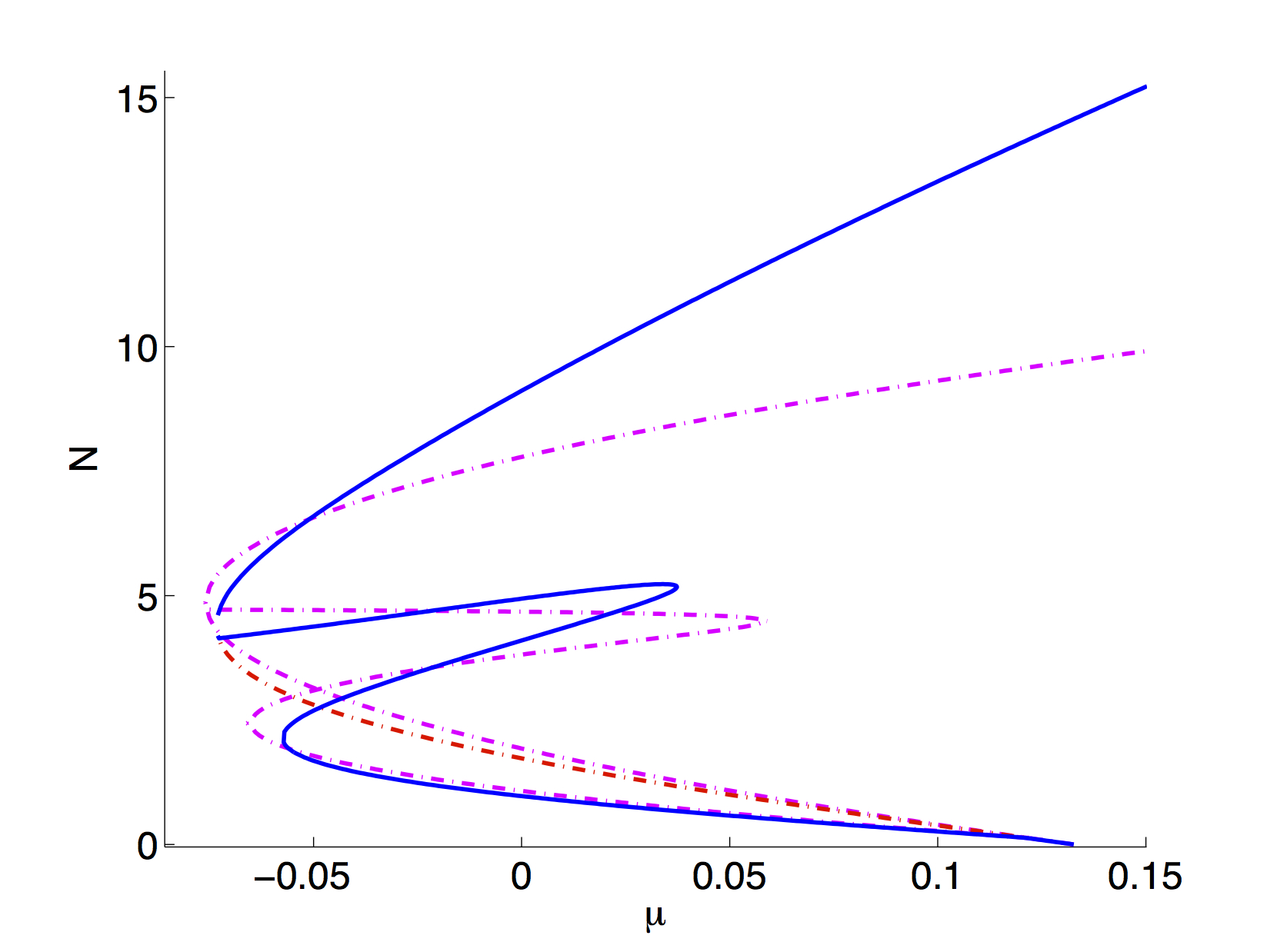}
\includegraphics[width = 7cm]{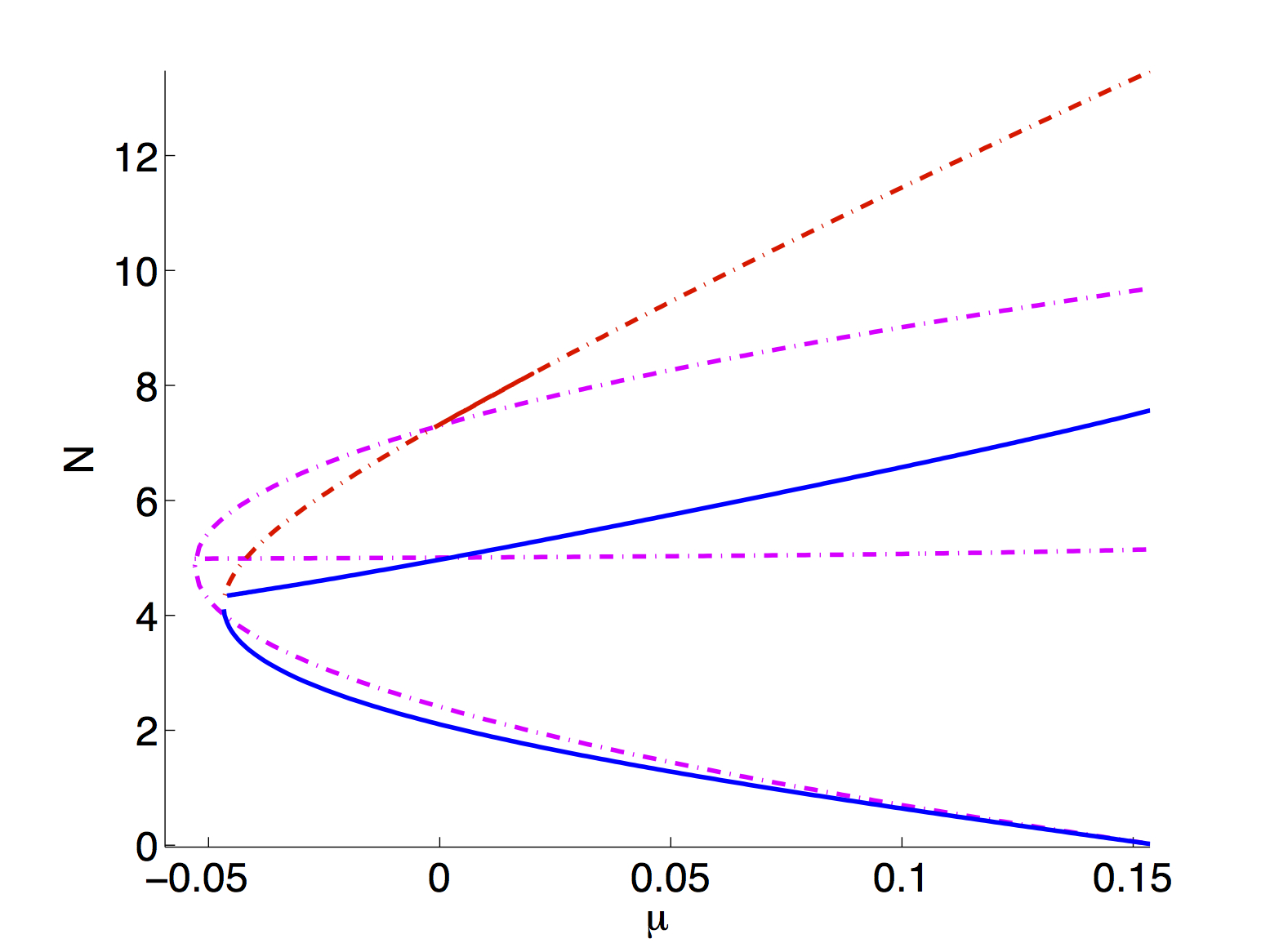}\\
\includegraphics[width = 7.5cm]{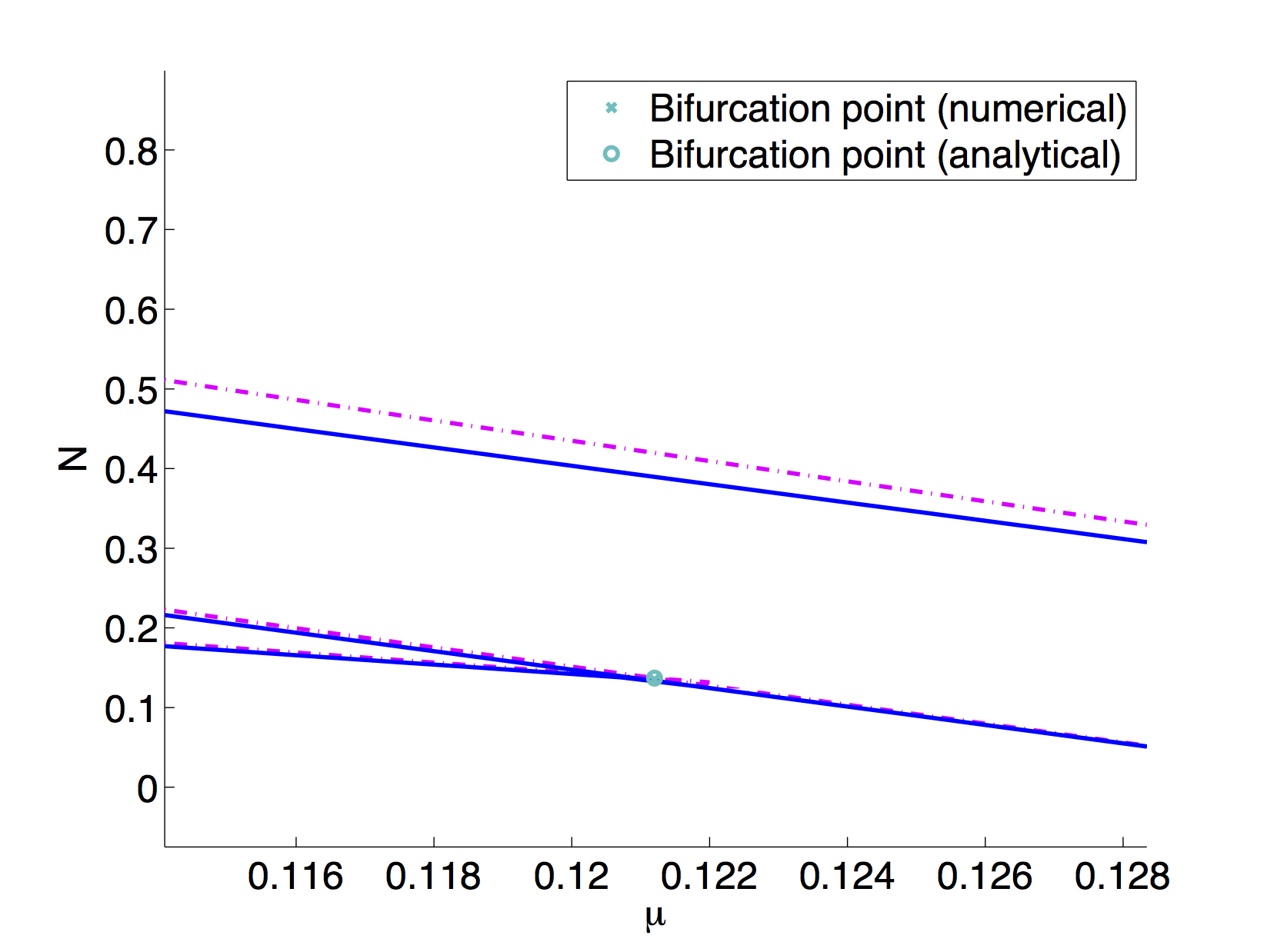}
\end{center}
\caption{Same as the previous three figures, but now for
the {\it focusing cubic/defocusing quintic} case of $s=-1$ and
$\delta=1$, for an intermediate interaction range of $\sigma=1$.}
\label{stat3a}
\end{figure}

Finally, we examine also one case where we switch the signs of the nonlocal terms to $(s,\delta) = (-1,1)$, so now the cubic term is the one that behaves attractively while the quintic one behaves repulsively.
This is illustrated in Fig.~\ref{stat3a}.
The interaction range $\sigma$ is selected here to be $1$ and here we see that the same phenomenology appears in a region where the cheminal potential varies from $-0.08$ to $0.155$, thus attaining negative values. As earlier, both states emanate for the same values of $\mu$ and as we decrease its value we observe the symmetry breaking at $\mu = 0.1212$ (both for numerical and analytical) this time on the symmetric state which becomes unstable. As we further decrease
the chemical potential to negative values of $\mu$,
the symmetry restoring of the asymmetric state towards its parent symmetric
branch occurs at $\mu = -0.0727$ (numerical value). The analytical
prediction for this critical point is $\mu = -0.0755$.
Hence, once again we observe a good qualitative agreement for larger $N$
(although once again slight quantitative disparities exist between the
overall curves and the critical points in terms of $N$).
A look at the antisymmetric branch now shows us that a bifurcation occurs at the point where the solution changes slope ($dN/d\mu$), precisely at
$\mu = -0.0465$ (numerical) and is theoretically predicted to arise
at $\mu = -0.0526$ (analytical) with the antisymmetric branch becoming
unstable past this critical point. Once again the zoom of the bottom panel
confirms the quantitative nature of the analytical-numerical agreement
for small values of $N$, which retains its qualitative value even for
larger $N$.

\subsection{Dynamics}

\begin{figure}
\begin{center}
\includegraphics[width = 8cm]{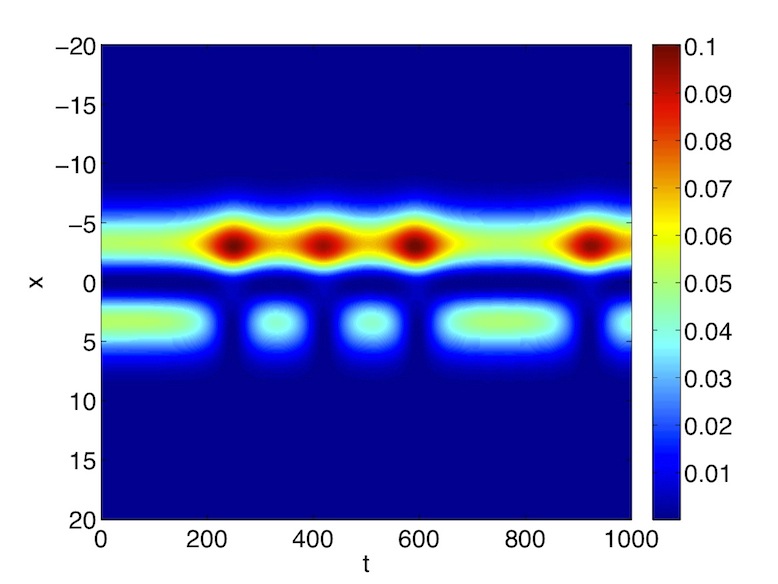}
\includegraphics[width = 8cm]{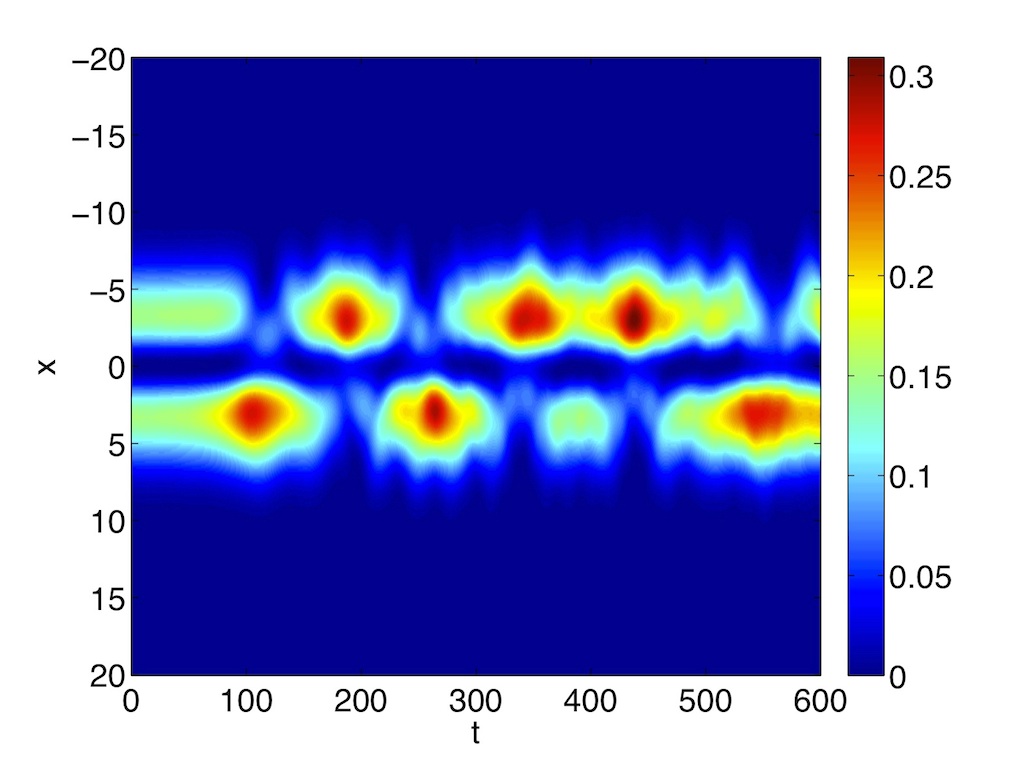}
\end{center}
\caption{Spatio-temporal contour plot of the density of the unstable solutions
when $\sigma = 1$, for $s=1$ and $\delta=-1$. The panels are initialized
with (a weakly perturbed case example of)
the antisymmetric solution for $\mu = 0.19$ and $0.25$ (left and
right, respectively).}
\label{evolution}
\end{figure}

\begin{figure}
\begin{center}
\includegraphics[width = 10cm]{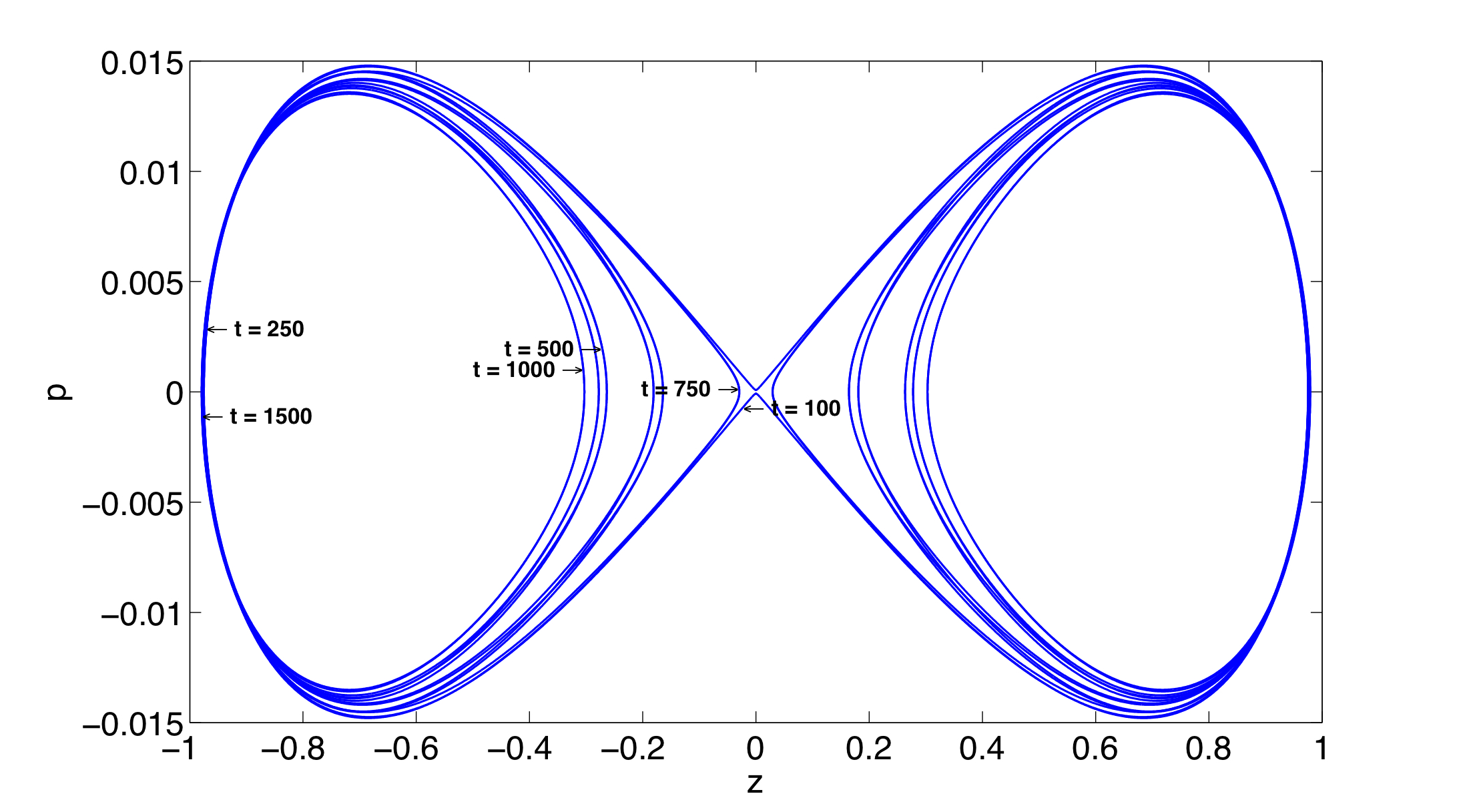}
\includegraphics[width = 12cm]{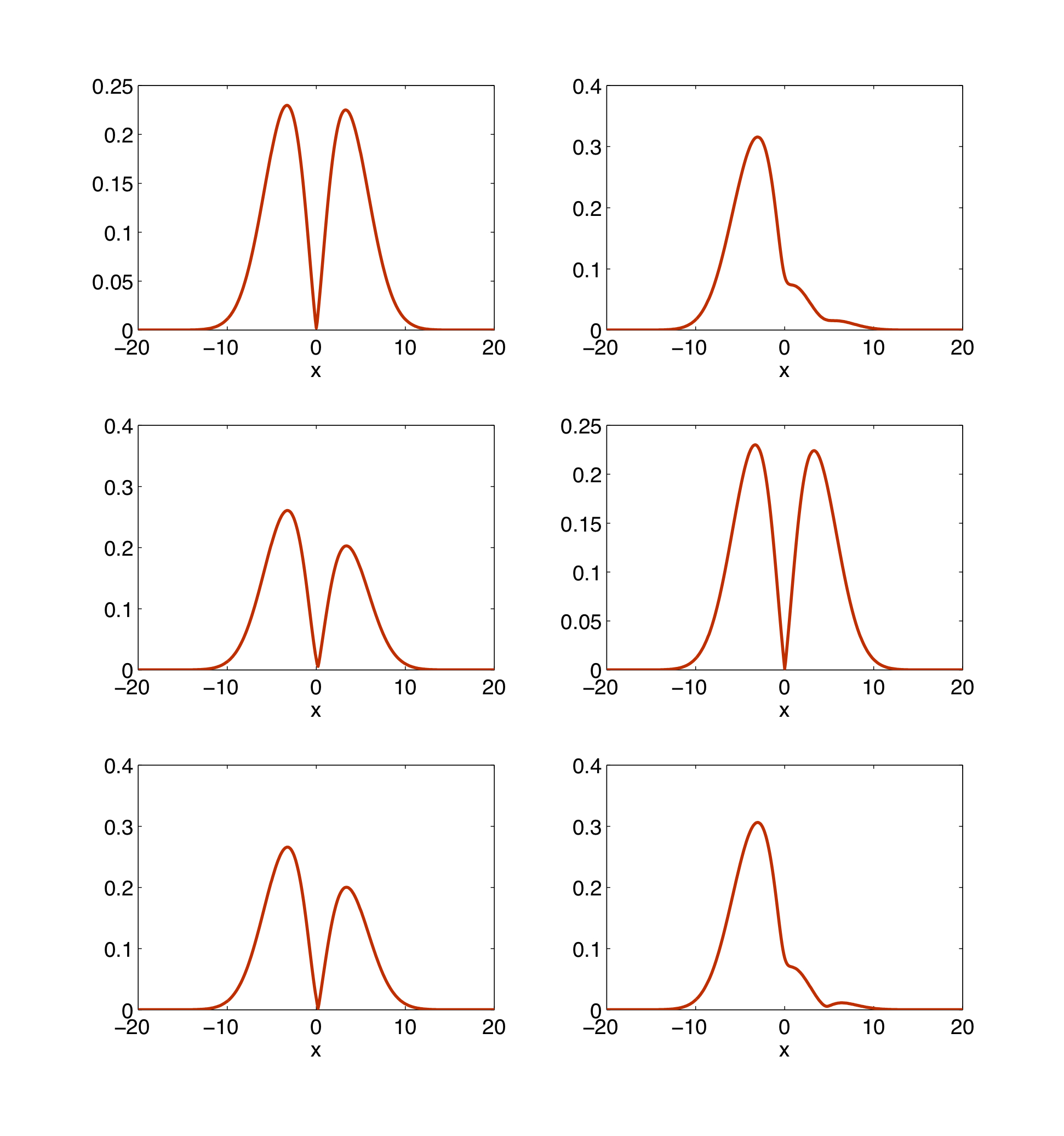}
\end{center}
\caption{Top panel: the numerically obtained trajectory of the solution for
$\mu = 0.19$, for times between $0$ and $1500$. Rows below: the profiles of
the solution for $t = 100$, $250$ (second row),
$500$, $750$ (third row), $1000$ and $1500$ (fourth row).}
\label{pde_phase_plane}
\end{figure}

Finally, we briefly turn to the dynamics of the system, in order
to observe the implications of the dynamical instability
due to the symmetry breaking. The relevant evolution of
the unstable solutions for $\mu=0.19$ and $\mu=0.25$, in the
case of $\sigma=1$ (recall that $s=1$ and $\delta=-1$) are shown
in Fig.~\ref{evolution}. In both cases, it can be seen that
the weak perturbation added on top of the exact numerical
solution in the initial conditions has a projection along the
unstable eigenmode. This projection, for sufficiently long times
(about $200$ in the left panel and about $100$ in the right panel),
gets amplified and eventually leads to a visible (i.e., of order
unity) symmetry breaking in the profile of the state. While the
space-time evolution of the density (in the atomic case; optical
intensity in the optical case) is shown in Fig.~\ref{evolution},
an interesting alternative way to visualize the instability was proposed
recently by~\cite{marzuola}. In the latter work, the PDE dynamics
was, in fact, projected to the phase plane of the two-mode approximation
and visualized therein. An example of such a visualization for the
case of $\mu=0.19$ can be seen in Fig.~\ref{pde_phase_plane}.
From both the phase plane curves and the profiles illustrated
underneath of the solution at different times, we can extract
some interesting conclusions. In particular, in the one degree of
freedom reduction of our theoretical analysis, the trajectory occurs
over iso-contours of the energy. Hence, the kind of phase plane
picture shown in Fig.~\ref{pde_phase_plane} would only be possible
by ``conglomerating'' many distinct orbits. However, it is important
to appreciate that the PDE has infinitely many degrees of freedom.
In that capacity, it is possible for the ``subspace'' of our two-mode
approximation to {\it dissipate} energy  towards
(or possibly regain energy from) higher energy
states (of the point spectrum of the system). In so doing, it appears
as if the system visits further and further inward trajectories of
lower energy, because indeed the excess energy has been imparted
to other degrees of freedom. This yields a clear illustration of
how the subspace of our two-modes is a {\it closed system} for
the ODE reduction, but instead is an {\it open system} for the
full PDE evolutionary dynamics~\footnote{It should be highlighted
that this process is not uni-directional. More specifically, the harmonically
trapped nature of the system may contribute to the reversal of the
above described type of ``flow'', leading to the eventual revisiting of
outward trajectories.}.

\section{Conclusions}

In the present work, we examined double well potentials in the presence of
nonlocal interactions both in the cubic and in the quintic part of the
nonlinearity. We attempted to address such settings by means of a two-mode
decomposition that has the notable advantage that nonlocality is
not substantially different to handle therein, as the nonlocal kernels
merely contribute to relevant overlap integrals that need some systematic
book-keeping, but are otherwise not considerably harder than is the locally
nonlinear case. There are some particularly important attributes
of the quintic case
that we were able to extract via a normal form reduction and phase
plane visualization
(under suitable circumstances
of ``competition'' e.g. for a defocusing cubic but focusing quintic
nonlinearity). One such is that contrary to the purely cubic case,
the reduction is able to predict not only a symmetry breaking
bifurcation, but {\it also} a symmetry restoring one (at least
for a suitable interval of range parameters for the interaction
kernel). Another unusual characteristic is that symmetry breaking
bifurcations are encountered {\it both} for the symmetric and
the antisymmetric branch, again differently than is the case for the
cubic nonlinearity in the double well setting. These features were
tested against numerical bifurcation results and good agreement
was found where appropriate (e.g. low atom numbers and a suitable range of
the interaction range). Disparities arising for high $N$ and
large $\sigma$ were systematically explained. Finally, the
instability dynamics was visualized not only by space-time density
evolution plots but also by offering its projection
to the phase plane of the double well theoretical reduction
and assessing the similarities and differences therein of the
ODE approximation and full PDE result.

There are numerous possibilities for the extension of the present
results to more elaborate contexts. On the one hand, even in the
one-dimensional setting, one could envision a study of different
interaction ranges between the cubic and quintic terms (or, for that
matter, combinations of local and nonlocal nonlinearities within the
cubic and/or quintic terms). On the other hand, extensions to one
dimensional settings with more wells would bring along a richer
phenomenology (in that setting the three-well local case has been
studied~\cite{tk} and was recently revisited in~\cite{good}), while
in higher dimensional settings such as 2d, four well settings in a square
configuration~\cite{four} or other configurations exploiting the
geometry of the system would be interesting to study.


\acknowledgments PGK gratefully acknowledges support from the
National Science Foundation under grants DMS-0806762 and
CMMI-1000337, as well as by the Alexander von Humboldt Foundation
through a Research Fellowship, the Alexander S. Onassis Public
Benefit Foundation (grant RZG 003/2010-2011) and the Binational
Science Foundation (grant 2010239). The work of PAT was partially
supported by the State Scholarships Foundation in Greece. VMR
gratefully acknowledges support from Research Council of AUTH
(Grant 87872). This research has been co-financed by the European
Union (European Social Fund - ESF) and Greek national funds
through the Operational Program "Education and Lifelong Learning"
of the National Strategic Reference Framework (NSRF) - Research
Funding Program: THALES. Investing in knowledge society through
the European Social Fund.

\vspace{5mm}

\appendix{{\bf Justification of the Dynamical Equation in Nonlinear Optics}}

The standard 1D model of the thermal optical nonlinearity is based on the
following system\ (see, e.g., Refs. \cite{A,boris_add2}):%
\begin{gather}
iu_{z}+\frac{1}{2}u_{xx}+mu=0,  \label{u} \\
m-dm_{xx}=\sigma _{0}\left\vert u\right\vert ^{2},  \label{m}
\end{gather}%
where $d>0$ is the squared correlation length of the nonlocal
nonlinearity, the real field $m$ is a local perturbation of the
refraction index, and $\sigma _{0}$ is the coefficient of the
optical absorption which leads to heating of the medium, so that
$\sigma _{0}\left\vert u\right\vert ^{2}$ is the local source in the
effective heat-conductivity equation (\ref{m}). If the heating is
provided by the resonant absorption by dopants, the absorption may
be saturable. The saturation may be described,
in the
simplest approximation, by the following modification of Eq. (\ref{m}):%
\begin{equation}
m-dm_{xx}=\sigma _{0}\left\vert u\right\vert ^{2}-\sigma _{0}|u|^{4}.
\label{02}
\end{equation}

Finally, an obvious solution of Eq. (\ref{02}) is
\begin{equation*}
m(x)=\frac{\sqrt{d}}{2}\int_{-\infty }^{+\infty }\exp \left( -\frac{1}{\sqrt{%
d}}\left\vert x-x^{\prime }\right\vert \right) \left[ \sigma _{0}\left\vert
u(x^{\prime })\right\vert ^{2}-\sigma _{0}|u(x^{\prime })|^{4}\right]
dx^{\prime }.
\end{equation*}%
The substitution of this into Eq. (\ref{m}) leads to the nonlocal NLS
equation with the cubic-quintic nonlinearity.


\begin{thebibliography}{99}

\bibitem{sulem} C. Sulem and P. L. Sulem,
\newblock  {\it The Nonlinear
Schr{\"o}dinger Equation} (Springer-Verlag, New York, 1999).

\bibitem{ablowitz} M.J. Ablowitz, B. Prinari and A.D. Trubatch,
{\it Discrete and Continuous Nonlinear Schr{\"o}dinger Systems},
Cambridge University Press (Cambridge, 2004).


\bibitem{markus1} M. Albiez, R. Gati, J. F\"{o}lling, S. Hunsmann, M.
Cristiani, and M. K. Oberthaler, Phys. Rev. Lett. \textbf{95}, 010402 (2005).


\bibitem{markus2} T. Zibold, E. Nicklas, C. Gross and
M.K. Oberthaler, Phys. Rev. Lett. {\bf 105}, 204101 (2010).

\bibitem{smerzi} S. Raghavan, A. Smerzi, S. Fantoni, and S. R. Shenoy, Phys.
Rev. A \textbf{59}, 620 (1999); S. Raghavan, A. Smerzi, and V. M. Kenkre,
Phys. Rev. A \textbf{60}, R1787 (1999); A. Smerzi and S. Raghavan, Phys.
Rev. A \textbf{61}, 063601 (2000).

\bibitem{kiv2} E. A. Ostrovskaya, Yu. S. Kivshar, M. Lisak, B. Hall, F.
Cattani, and D. Anderson, Phys. Rev. A \textbf{61}, 031601(R) (2000).

\bibitem{mahmud} K. W. Mahmud, J. N. Kutz, and W. P. Reinhardt, Phys. Rev. A
\textbf{66}, 063607 (2002).

\bibitem{bam} V. S. Shchesnovich, B. A. Malomed, and R. A. Kraenkel, Physica
D \textbf{188}, 213 (2004).

\bibitem{Bergeman_2mode} D. Ananikian and T. Bergeman, Phys. Rev. A \textbf{%
73}, 013604 (2006).

\bibitem{infeld} P. Zi\'{n}, E. Infeld, M. Matuszewski, G. Rowlands, and M.
Trippenbach, Phys. Rev. A \textbf{73}, 022105 (2006).

\bibitem{todd} T. Kapitula and P. G. Kevrekidis, Nonlinearity \textbf{18},
2491 (2005).

\bibitem{theo} G. Theocharis, P. G. Kevrekidis, D. J. Frantzeskakis, and P.
Schmelcher, Phys. Rev. E \textbf{74}, 056608 (2006).

\bibitem{carr} D. R. Dounas-Frazer, A. M. Hermundstad, and L. D. Carr, Phys.
Rev. Lett. \textbf{99}, 200402 (2007).

\bibitem{pseudo} T. Mayteevarunyoo, B. A. Malomed, and G. Dong. Phys. Rev. A
\textbf{78}, 053601 (2008).

\bibitem{fibers} C. Par\'{e} and M. Florja\'{n}czyk, Phys. Rev. A \textbf{41}%
, 6287 (1990); A. I. Maimistov, Kvant. Elektron. \textbf{18}, 758 (1991) [In
Russian; English translation: Sov. J. Quantum Electron. \textbf{21}, 687; W.
Snyder, D. J. Mitchell, L. Poladian, D. R. Rowland, and Y. Chen, J. Opt.
Soc. Am. B \textbf{8}, 2102 (1991); P. L. Chu, B. A. Malomed, and G. D.
Peng, J. Opt. Soc. Am. B \textbf{10}, 1379 (1993); N. Akhmediev, and A.
Ankiewicz, Phys. Rev. Lett. \textbf{70}, 2395 (1993); B. A. Malomed, I.
Skinner, P. L. Chu, and G. D. Peng, Phys. Rev. E \textbf{53}, 4084 (1996).

\bibitem{HaeltermannPRL02} C. Cambournac, T. Sylvestre, H. Maillotte , B.
Vanderlinden, P. Kockaert, Ph. Emplit, and M. Haelterman, Phys. Rev. Lett.
\textbf{89}, 083901 (2002).

\bibitem{zhigang} P. G. Kevrekidis, Z. Chen, B. A. Malomed, D. J.
Frantzeskakis, and M. I. Weinstein, Phys. Lett. A \textbf{340}, 275 (2005).


\bibitem{Cr} A. Griesmaier, J. Werner, S. Hensler, J. Stuhler, and T. Pfau,
Phys. Rev. Lett. \textbf{94}, 160401 (2005); J. Stuhler, A. Griesmaier, T.
Koch, M. Fattori, T. Pfau, S. Giovanazzi, P. Pedri, and L. Santos, \textit{%
ibid}. \textbf{95}, 150406 (2005); J. Werner, A. Griesmaier, S. Hensler, J.
Stuhler, and T. Pfau, \textit{ibid}. \textbf{94}, 183201 (2005); A.
Griesmaier, J. Stuhler, T. Koch, M. Fattori, T. Pfau, and S. Giovanazzi,
\textit{ibid}. \textbf{97}, 250402 (2006); A. Griesmaier, J. Phys. B: At.
Mol. Opt. Phys. \textbf{40}, R91 (2007); T. Lahaye, T. Koch, B. Fr\"{o}%
hlich, M. Fattori, J. Metz, A. Griesmaier, S. Giovanazzi, and T. Pfau,
Nature (London) \textbf{448}, 672 (2007).

\bibitem{review} T. Lahaye, C. Menotti, L. Santos, M. Lewenstein and T.
Pfau, Rep. Progr. Phys. \textbf{72}, 126401 (2009).

\bibitem{pra09} B. Xiong, J. Gong, H. Pu, W. Bao, and B. Li, Phys. Rev. A
\textbf{79}, 013626 (2009), M. Asad-uz-Zaman and D. Blume, \textit{ibid}.
\textbf{80}, 053622 (2009).


\bibitem{hetmol} T. K\"{o}hler, K. G\'{o}ral, and P. S. Julienne, Rev. Mod.
Phys. \textbf{78}, 1311 (2006); J. Sage, S. Sainis, T. Bergeman, and D.
DeMille, Phys. Rev. Lett. \textbf{94}, 203001 (2005); C. Ospelkaus, L.
Humbert, P. Ernst, K. Sengstock, and K. Bongs, \textit{ibid}. \textbf{97},
120402 (2006); J. Deiglmayr, A. Grochola, M. Repp, K. M\"{o}rtlbauer, C. Gl%
\"{u}ck, J. Lange, O. Dulieu, R. Wester, and M. Weidem\"{u}ller, \textit{ibid%
}. \textbf{101}, 133004 (2008); F. Lang, K. Winkler, C. Strauss, R. Grimm,
and J. H. Denschlag, \textit{ibid}. \textbf{101}, 133005 (2008).

\bibitem{dc} M. Marinescu and L. You, Phys. Rev. Lett. \textbf{81}, 4596
(1998); S. Giovanazzi, D. O'Dell, and G. Kurizki, Phys. Rev. Lett. \textbf{88%
}, 130402 (2002); I. E. Mazets, D. H. J. O'Dell, G. Kurizki, N. Davidson,
and W. P. Schleich, J. Phys. B 37, S155 (2004); R. L\"{o}w, R. Gati, J.
Stuhler and T. Pfau, Europhys. Lett. \textbf{71}, 214 (2005).

\bibitem{krol1} W. Kr\'{o}likowski, O. Bang, J. J. Rasmussen, and J. Wyller,
Phys. Rev. E \textbf{64}, 016612 (2001); O. Bang, W. Kr\'{o}likowski, J.
Wyller and J. J. Rasmussen, Phys. Rev. E \textbf{66}, 046619 (2002); J.
Wyller, W. Kr\'{o}likowski, O. Bang and J. J. Rasmussen, Phys. Rev. E
\textbf{66}, 066615 (2002).

\bibitem{A}  W. Krolikowski, O. Bang, N.I. Nikolov, D. Neshev, J. Wyller, J.J.
Rasmussen, and D. Edmundson, 
J. Opt. B 
{\bf 6}, S288 (2004).


\bibitem{krolik} D. Briedis, D. E. Petersen, D. Edmundson, W. Kr\'{o}%
likowski, and O. Bang, Opt. Exp. \textbf{13}, 435 (2005).

\bibitem{moti1} C. Rotschild, O. Cohen, O. Manela, and M. Segev, Phys. Rev.
Lett. \textbf{95}, 213904 (2005).

\bibitem{krolik2} A. Dreischuh, D.N. Neshev, D.E. Petersen, O. Bang, and
W. Krolikowski
Phys. Rev. Lett. {\bf 96}, 043901 (2006).

\bibitem{B} N.I. Nikolov, D. Neshev, O. Bang, W.Z. Krolikowski, Phys. Rev. E \textbf{68}, 036614
(2003).

\bibitem{C} P.V. Larsen, M.P. Sorensen, O. Bang, W.Z. Krolikowski, S.
Trillo,  Phys. Rev. E \textbf{73}, 036614 (2006).

\bibitem{D} M. Bache, O. Bang, J. Moses, F.W. Wise, Opt. Lett. \textbf{32}, 2490 (2007)

\bibitem{E} M. Bache, O. Bang, W. Krolikowski, J. Moses, F.W. Wise,  Opt.
Express \textbf{16}, 3273 (2008).



\bibitem{F} J.F. Corney, O. Bang, Phys. Rev. E \textbf{64}, 047601 (2001)

\bibitem{PTS}
B.~L. Lawrence and G.~I. Stegeman.
Two-dimensional bright spatial solitons stable over limited
intensities and ring formation in polydiacetylene para-toluene sulfonate.
{\em Optics letters}, {{\bf 23}}, 8 (1998) 591--593.

\bibitem{CQglass}
F. Smektala, C. Quemard, V. Couderc, and A. Barth\'{e}l\'{e}my,
J. Non-Cryst. Solids 274, 232 (2000);
G. Boudebs, S. Cherukulappurath, H. Leblond, J. Troles,
F. Smektala, and F. Sanchez, Opt. Commun. 219, 427 (2003).

\bibitem{CQorganic}
C. Zhan et al., D. Zhang, D. Zhu, D. Wang, Y. Li, D. Li, Z. Lu,
L. Zhao, and Y. Nie,
J. Opt. Soc. Am. B 19, 369 (2002).

\bibitem{colloid}
G. S. Agarwal and S. Dutta Gupta, Phys. Rev. A 38, 5678 (1988);
E. L. Falc\~{a}o-Filho, C. B. de Ara\'{u}jo, and J. J. Rodrigues, Jr,
J. Opt. Soc. Am. B 24, 2948 (2007).

\bibitem{dye}
R. A. Ganeev et al., M. Baba, M. Morita, A. I. Ryasnyansky, M. Suzuki,
M. Turu, H. Kuroda, J. Opt. A: Pure Appl. Opt. 6, 282 (2004).

\bibitem{ferroelectric}
B. Gu, Y. Wang, W. Ji, and J. Wa, Appl. Phys. Lett. 95, 041114 (2009).

\bibitem{cascading}
K. Dolgaleva, H. Shin, and R. W. Boyd, Phys. Rev. Lett. 103, 113902 (2009).

\bibitem{chenyu} C. Wang, P.G. Kevrekidis, D.J. Frantzeskakis and B.A.
Malomed, Physica D {\bf 240}, 805 (2011).

\bibitem{jy} J. Yang,
Stud. Appl. Math. {\bf 129}, 133 (2012);
Physica D {\bf 244}, 50 (2013).

\bibitem{boris_add} B. B. Baizakov, F. Kh. Abdullaev, B. A. Malomed, and 
M. Salerno,
J. Phys. B: At. Mol. Opt. Phys. {\bf 42}, 175302 (2009).


\bibitem{boris_add2} Z. Xu, Y. V.
Kartashov, and L. Torner, Opt. Lett. \textbf{30}, 317 (2005)

\bibitem{engels} P. Engels and C. Atherton
Phys. Rev. Lett. {\bf 99}, 160405 (2007).

\bibitem{jyang} J. Yang, Phys. Rev. E {\bf 85}, 037602 (2012).

\bibitem{marzuola} J. Marzuola and M.I. Weinstein,
Discr. Cont. Dyn. Sys. A {\bf 28}, 1505 (2010).

\bibitem{tk} T. Kapitula, P.G. Kevrekidis and Z. Chen,
SIAM J. Appl. Dyn. Sys. {\bf 5}, 598 (2006).

\bibitem{good} R. Goodman,
J. Phys. A {\bf 44}, 425101 (2011).

\bibitem{four} C. Wang, G. Theocharis, P.G. Kevrekidis,
N. Whitaker, K.J.H. Law, D.J. Frantzeskakis and B.A. Malomed,
Phys. Rev. E {\bf 80}, 046611 (2009).

\end{thebibliography}
\end{document}